\documentclass{appolb}
\usepackage{bbm}
\usepackage[final]{graphics}
\usepackage{amsmath}
\usepackage{amsfonts,amsbsy}
\usepackage{amssymb}
\usepackage[dvips]{graphicx}

\def\empile#1\over#2{\mathrel{\mathop{\kern 0pt#1}\limits_{#2}}}
\def\bs{\boldsymbol}

\newcommand{\slv}{\raise.15ex\hbox{$/$}\kern-.53em\hbox{$v$}}
\newcommand{\slF}{\raise.15ex\hbox{$/$}\kern-.53em\hbox{$F$}}
\newcommand{\slL}{\raise.15ex\hbox{$/$}\kern-.53em\hbox{$L$}}
\newcommand{\slP}{\raise.15ex\hbox{$/$}\kern-.53em\hbox{$P$}}
\newcommand{\slp}{\raise.15ex\hbox{$/$}\kern-.53em\hbox{$p$}}
\newcommand{\slq}{\raise.15ex\hbox{$/$}\kern-.53em\hbox{$q$}}
\newcommand{\slR}{\raise.15ex\hbox{$/$}\kern-.53em\hbox{$R$}}
\newcommand{\slQ}{\raise.15ex\hbox{$/$}\kern-.53em\hbox{$Q$}}
\newcommand{\slK}{\raise.15ex\hbox{$/$}\kern-.53em\hbox{$K$}}
\newcommand{\slk}{\raise.15ex\hbox{$/$}\kern-.53em\hbox{$k$}}
\newcommand{\slD}{\raise.15ex\hbox{$/$}\kern-.53em\hbox{$D$}}
\newcommand{\slC}{\raise.15ex\hbox{$/$}\kern-.53em\hbox{$C$}}
\newcommand{\slA}{\raise.15ex\hbox{$/$}\kern-.53em\hbox{$A$}}
\newcommand{\slSigma}{\raise.15ex\hbox{$/$}\kern-.53em\hbox{$\Sigma$}}
\newcommand{\slpartial}{\raise.15ex\hbox{$/$}\kern-.53em\hbox{$\partial$}}
\newcommand{\slcalP}{\raise.15ex\hbox{$/$}\kern-.63em\hbox{$\cal P$}}

\def\p{{\boldsymbol p}}
\def\q{{\boldsymbol q}}
\def\k{{\boldsymbol k}}

\def\x{{\boldsymbol x}}
\def\y{{\boldsymbol y}}

\def\r{{\boldsymbol r}}

\def\b{{\boldsymbol b}}

\catcode`\@=11

\newcount\@tempcntc
\def\@citex[#1]#2{\if@filesw\immediate\write\@auxout{\string\citation{#2}}\fi
  \@tempcnta\z@\@tempcntb\m@ne\def\@citea{}\@cite{%
        \@for\@citeb:=#2\do%
    {\@ifundefined{b@\@citeb}%
        {\@citeo\@tempcntb\m@ne\@citea%
                \def\@citea{,\penalty\@m\ }{\bf ?}\@warning%
                {Citation `\@citeb' on page \thepage \space undefined}}%
        {\setbox\z@\hbox{\global\@tempcntc0\csname b@\@citeb\endcsname\relax}%%
     \ifnum\@tempcntc=\z@ \@citeo\@tempcntb\m@ne%
       \@citea\def\@citea{,\penalty\@m}%
       \hbox{\csname b@\@citeb\endcsname}%
     \else%
      \advance\@tempcntb\@ne%
      \ifnum\@tempcntb=\@tempcntc%
      \else\advance\@tempcntb\m@ne\@citeo%
      \@tempcnta\@tempcntc\@tempcntb\@tempcntc\fi\fi}}\@citeo}{#1}}%

\def\@citeo{\ifnum\@tempcnta>\@tempcntb\else\@citea
  \def\@citea{,\penalty\@m}%
  \ifnum\@tempcnta=\@tempcntb\the\@tempcnta\else
   {\advance\@tempcnta\@ne\ifnum\@tempcnta=\@tempcntb \else
\def\@citea{--}\fi
    \advance\@tempcnta\m@ne\the\@tempcnta\@citea\the\@tempcntb}\fi\fi}

\catcode`\@=12

\begin{document}

\pagestyle{plain}
%% uncomment the following line to get equations numbered by (sec.num)
%\eqsec
\newcount\eLiNe\eLiNe=\inputlineno\advance\eLiNe by -1

\title{\bf Three lectures on multi-particle production in the Glasma}
\author{Fran\c cois Gelis$^{(1)}$, Raju Venugopalan$^{(2)}$}
\maketitle
\begin{center}
\begin{enumerate}
\item CEA, Service de Physique Th\'eorique (URA 2306 du CNRS),\\
  91191, Gif-sur-Yvette Cedex, France
\item Department of Physics, Bldg. 510 A,\\ Brookhaven National Laboratory,
  Upton, NY-11973, USA
\end{enumerate}
\vglue 5mm
\date{\today}
\end{center}

\begin{abstract}
  In the Color Glass Condensate (CGC) effective field theory, when two
  large sheets of Colored Glass collide, as in a central
  nucleus-nucleus collision, they form a strongly interacting,
  non-equilibrium state of matter called the Glasma. How Colored Glass
  shatters to form the Glasma, the properties of the Glasma, and the
  complex dynamics transforming the Glasma to a thermalized Quark
  Gluon Plasma (QGP) are questions of central interest in
  understanding the properties of the strongly interacting matter
  produced in heavy ion collisions. In the first of these lectures, we
  shall discuss how these questions may be addressed in the framework
  of particle production in a field theory with strong time dependent
  external sources. Albeit such field theories are non-perturbative
  even for arbitrarily weak coupling, moments of the multiplicity
  distribution can in principle be computed systematically in powers
  of the coupling constant. We will demonstrate that the average
  multiplicity can be (straightforwardly) computed to leading order in
  the coupling and (remarkably) to next-to-leading order as well. The
  latter are obtained from solutions of small fluctuation equations of
  motion with {\it retarded boundary conditions}. In the second
  lecture, we relate our formalism to results from previous 2+1 and
  3+1 dimensional numerical simulations of the Glasma fields. The
  latter show clearly that the expanding Glasma is unstable; small
  fluctuations in the initial conditions grow exponentially with the
  square root of the proper time. Whether this explosive growth of
  small fluctuations leads to early thermalization in heavy ion
  collisions requires at present a better understanding of these
  fluctuations on the light cone. In the third and final lecture,
  motivated by recent work of Bia\l{}as and
  Je\.zabek~\cite{BialasJezabek}, we will discuss how the widely
  observed phenomenon of limiting fragmentation is realized in the CGC
  framework.

\end{abstract}

\section*{Introduction}

The theme of these lectures at the 46th course of the Zakopane school
is multi-particle production in hadronic collisions at high energies
in the Color Glass Condensate (CGC) effective field theory.  There has
been tremendous progress in our theoretical understanding in the seven
years since one of the authors last lectured here. At that time, the
author's lectures covered the state of the art (in the CGC framework)
in both deeply inelastic scattering (DIS) studies and in hadronic
multi-particle production~\cite{Venugopalan:1999}.  A sign of rapid
progress in the field is that there were several talks and lectures at
this school covering various aspects of this physics in DIS alone. We
will restrict ourselves here to developments in our understanding of
multi-particle production in hadronic collisions. Another significant
development in the last seven years has been the large amount of data
from the Relativistic Heavy Ion Collider (RHIC) at BNL, key features
of which were nicely summarized in the white-papers of the experimental
collaborations~\cite{whitepapers} culminating in the announcement of
the discovery of a ``perfect fluid" at RHIC. The exciting experimental
observations were discussed here in the lectures of
Jacak~\cite{Jacak}. The RHIC data have had a tremendous impact on the
CGC studies of multi-particle production.  While we will discuss RHIC
phenomenology, and indeed specific applications of theory to data, our
primary focus will be on attempts to develop a systematic theoretical
framework in QCD to compute multi-particle production in hadronic
collisions.  Some applications of the CGC approach to RHIC
phenomenology were also covered at this school by Kharzeev as part of
his lectures~\cite{Kharzeev}. For recent comprehensive reviews, see
Ref.~\cite{BlaizotG:2005,Jalilian-MarianK:2005}.

The collider era in high energy physics has made possible
investigations of QCD structure at a deep level in studies of
multi-particle final states.  Much attention has been focused on the
nature of multi-particle production in jets;  for a nice review, see
Refs.~\cite{DremiG1,WolfDK1}. The problem is however very
general. Theoretical developments in the last couple of decades
suggest that semi-hard particle production in high energy hadronic
interactions is dominated by interactions between partons having a
small fraction $x$ of the longitudinal momentum of the incoming
nucleons.  In the Regge limit of small $x$ and fixed momentum transfer
squared $Q^2$ (corresponding to very large center of mass energies
$\sqrt{s}$) the Balitsky--Fadin--Kuraev--Lipatov (BFKL) evolution
equation~\cite{BalitL1,KuraeLF1} predicts that parton densities grow
very rapidly with decreasing $x$.  A rapid growth with $x$, in the gluon
distribution, for fixed $Q^2 \gg \Lambda_{_{\rm QCD}}^2$, was observed in
the HERA experiments~\cite{H1,ZEUS}. (It is not clear however that the
observed growth of the gluon distribution is a consequence of BFKL
dynamics~\cite{Forte}.) Because the rapid growth in the Regge limit
corresponds to very large phase space densities of partons in hadronic
wave functions, saturation effects may play an important role in
hadronic collisions at very high
energies~\cite{GriboLR1,MuellQ1,BlaizM1,Muell4}. These slow down the
growth of parton densities relative to that of BFKL evolution and may
provide the mechanism for the unitarization of cross-sections at high
energies.

The large parton phase space density suggests that small $x$ partons
can be described by a classical color field rather than as particles
\cite{McLerV1,McLerV2,McLerV3}. Light cone kinematics (more simply,
time dilation) further indicates that there is a natural separation in
time scales, whereby the small $x$ partons are the dynamical degrees
of freedom and the large $x$ partons act as static color sources for
the classical field.  In the McLerran--Venugopalan (MV) model, the
large color charge density of sources is given by the density of large
$x$ partons in a big nucleus which contains $3A$ valence quarks (where
$A$ is the atomic number of the nucleus). In this limit of strong
color sources, one has to solve the non--linear classical Yang-Mills
equations to obtain the classical field corresponding to the small $x$
parton modes. This procedure properly incorporates, at tree level, the
recombination interactions that are responsible for gluon
saturation. In the MV model, the distribution of large $x$ color
sources is described by a Gaussian statistical
distribution~\cite{McLerV1,Kovch1}. A more general form of this
statistical distribution, for $SU(N_c)$ gauge theories, valid for
large $A$ and moderate $x$, is given in Refs.~\cite{JeonV1,JeonV2}.

The separation between large $x$ and small $x$, albeit natural, is
somewhat arbitrary in the MV model; the physics should in fact be
independent of this separation of scales. This property was exploited
to derive a functional renormalization group (RG) equation, the JIMWLK
equation, describing the evolution of the gauge invariant source
distributions to small
$x$~\cite{JalilKMW1,JalilKLW1,JalilKLW2,JalilKLW3,JalilKLW4,IancuLM1,IancuLM2,FerreILM1}.
The JIMWLK functional RG equation is equivalent to an infinite
hierarchy of evolution equations describing the behavior of
multi-parton correlations at high energies first derived by
Balitsky~\cite{Balit1}. A useful (and tremendously simpler) large
$N_c$ and large $A$ mean-field approximation independently derived by
Kovchegov~\cite{Kovch3}, is commonly known as the Balitsky-Kovchegov
equation. The general effective field theory framework describing the
non-trivial behavior of multi-parton correlations at high energies is
often referred to as the Color Glass Condensate
(CGC)~\cite{McLer1,IancuLM3,IancuV1}.

Several lecturers at the school discussed the state of the art in our
understanding of the small $x$
wavefunction~\cite{McLerran}. For previous discussions at
recent schools, see Refs.~\cite{Dionysis,Anna,Kovner}.  To compute
particle production in the CGC framework, in addition to knowing the
distribution of sources in the small $x$ nuclear wavefunction, one must
calculate the properties of multi-particle production for any
particular configuration of sources. In this paper, we will assume
that the former is known. All we require is that these sources (as the
RG equations tell us) are strong sources, parametrically of the order
of the inverse coupling constant, and are strongly time dependent. We
will describe a formalism to compute multi-particle production for an
arbitrary distribution of such sources~\footnote{That one can separate
the properties of partons in the wave function from those in the final
state is a statement of factorization. This has not yet been
proven. We will briefly describe work in that direction in these
lectures.}.

These lectures are organized as follows. In the first lecture, we shall 
describe the formalism for computing particle production in a field
theory coupled to strong time-dependent external classical sources.
We will  consider as a toy model a $\phi^3$ scalar theory, where
$\phi$ is coupled to a strong external source. Although the
complications of QCD -- such as gauge invariance -- are very
important, many of the lessons gained from this simpler scalar theory
apply to studies of particle production in QCD.  We will demonstrate
that there is no simple power counting in the coupling constant $g$
for the probability $P_n$ to produce $n$ particles. A simple power
counting however exists for moments of $P_n$. We will discuss how one
computes the average multiplicity and (briefly) the variance. With
regard to the former, we will show how it can be computed to
next-to-leading order in the multiplicity. We will also discuss what
it takes to compute the generating function for the multiplicity
distribution to leading order in the coupling.

In lecture II, we will relate the formal considerations developed in
lecture I, to the results of real time numerical simulations of the
average multiplicity of gluons and quarks produced in heavy ion
collisions. Inclusive gluon production, to lowest order in the loop
expansion discussed in lecture I, is obtained by solving the classical
Yang-Mills equations for two color sources moving at the speed of
light in opposite directions~\cite{KovneMW1,KovneMW2,KovchR1}. This
problem has been solved numerically
in~\cite{KrasnV4,KrasnV1,KrasnV2,KrasnNV1,KrasnNV2,Lappi1} for the
boost-invariant case. The multiplicity of quark-pairs is computed from
the quark propagator in the background field of
\cite{KrasnV4,KrasnV1,KrasnV2,KrasnNV1,KrasnNV2,Lappi1} -- it has been
studied numerically in \cite{GelisKL1,GelisKL2}. A first computation
for the boost non-invariant case has also been performed
recently~\cite{PaulRaju1,PaulRaju2}.  It was shown there that rapidity
dependent fluctuations of the classical fields lead to the
non--Abelian analog of the Weibel instability~\cite{Weibel}, first
studied in the context of electromagnetic plasmas.  Such instabilities
may be responsible for the early thermalization required by
phenomenological studies of heavy ion collisions.  We will discuss how
a better understanding of the small quantum fluctuations discussed in
lecture I may provide insight into early thermalization.

In the third and final lecture, we will discuss how the formalism
outlined in lecture I simplifies in the case of proton-nucleus
collisions. (A similar simplification occurs in hadron-hadron
collisions at forward/backward rapidities where large $x$'s in one
hadron (small color charge density) and small $x$ in the other (large
color charge density) are probed.)  At leading order in the coupling,
lowest order in the proton charge density, and all orders in the
nuclear color charge density, analytical results are available for
both inclusive gluon and quark production~\footnote{For quark
  production, $k_\perp$ factorization breaks down even at leading
  order in pA collisions~\cite{BGV2}.}. In the former case, the
analytical formula can be written, in $k_\perp$ factorized form, as
the product of unintegrated distributions in both the proton and the
nucleus convoluted with the matrix element for the interactions
squared. This formula is used extensively in the literature for
phenomenological applications. We will discuss one such application,
that of limiting fragmentation.

\section*{Lecture I: How Colored Glass shatters to form the Glasma}

\begin{figure}[htbp]
\begin{center}
\resizebox*{8cm}{!}{\includegraphics{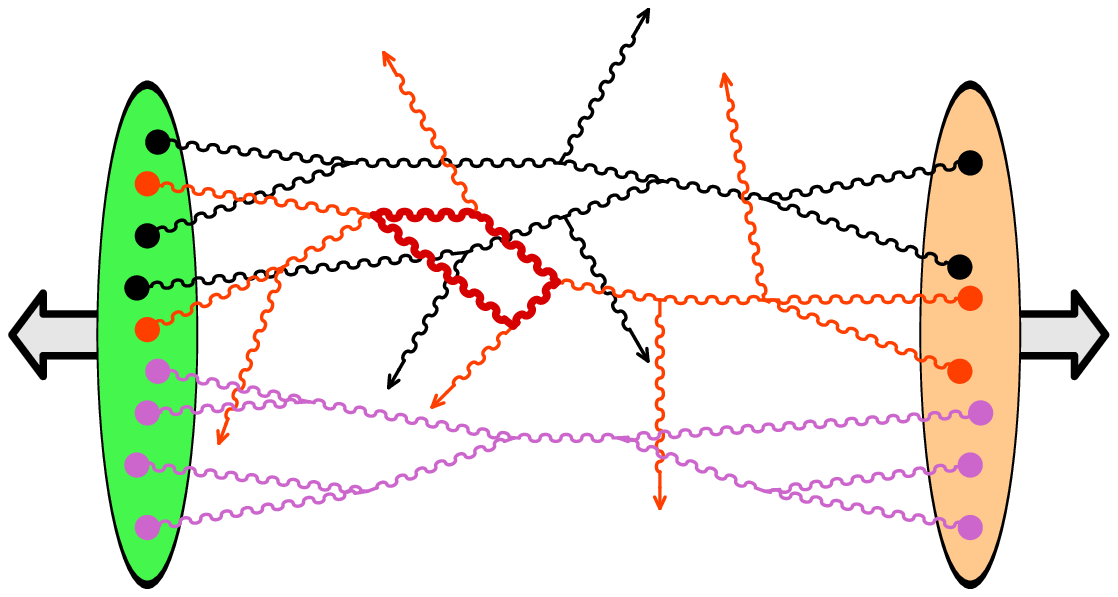}}
\end{center}
\caption{\label{fig:figI-1}Typical contribution to gluon production in
the collision of two sheets of Colored Glass. The dots denote the
color sources that describe the fast partons in the CGC framework.}
\end{figure}

As outlined in the introduction, the Glasma is formed when two sheets
of Colored Glass collide, producing a large number of partons. A
cartoon depicting this collision is shown in fig.~\ref{fig:figI-1}.
In the CGC framework, it is expected that observables can be expressed
as\footnote{As we shall see in lecture II, this formula may also
  require an average over some quantum fluctuations, because of an
  instability in boost invariant classical solutions of the Yang-Mills
  equations.}
\begin{equation}
\langle {\cal O}\rangle_{_Y} 
= \int [D\rho_1]\,[D\rho_2]\, 
W_{_{Y_{\rm beam}-Y}}[\rho_1]\,W_{_{Y_{\rm beam}+Y}}
[\rho_2] \, {\cal O}[\rho_1,\rho_2] \;,
\label{eq:LI-0}
\end{equation}
where they are first computed as a functional of the color charge
densities $\rho_1$ and $\rho_2$ of the two nuclei and then averaged
over all possible configurations of these sources, with the likelihood
of a particular configuration at a given rapidity $Y$ (= $\ln(1/x)$)
specified by the weight functionals $W_{_{Y_{\rm beam}-Y}}[\rho_1]$
and $W_{_{Y_{\rm beam}+Y}}[\rho_2]$ respectively. Here $Y_{\rm beam} =
\frac{1}{2} \ln(s/m_{\rm p}^2)$ is the beam rapidity in a hadronic collision
($m_{\rm p}$ denoting the proton mass) with the center of mass energy
$\sqrt{s}$.

The evolution of the $W_{_Y}[\rho]$'s with rapidity is described by
the JIMWLK
equation~\cite{JalilKMW1,JalilKLW1,JalilKLW2,JalilKLW3,JalilKLW4,IancuLM1,IancuLM2,FerreILM1}. For
small $x$ (large $Y$) and/or large nuclei, the rapid growth of parton
densities corresponds to light cone source densities
$\rho_1,\rho_2\sim 1/g$ -- in other words, the sources are
strong. Thus understanding how two sheets of Colored Glass shatter to
produce the Glasma requires that we understand the nature of particle
production in a field theory with strong, time dependent sources. In
this lecture, we will outline the tools to systematically compute
particle production in such theories. More details can be found in
Refs.~\cite{GelisV1,GelisV2}.

Field theories with strong time dependent sources are different from
field theories in the vacuum in one key respect. The ``vacuum'' in the
former, even in weak coupling, is non-trivial because it can produce
particles. Specifically, the amplitude from the vacuum state $|0_{\rm
in}\rangle$ to a populated state $|\alpha_{\rm out}\rangle$ is
\begin{equation}
\big<\alpha_{\rm out}\big|0_{\rm in}\big>\not=0\; .
\end{equation}
Unitarity requires that the sum over all ``out" states satisfies the identity
\begin{equation}
\sum_\alpha \Big|\big<\alpha_{\rm out}\big|0_{\rm in}\big>\Big|^2=1\; . 
\end{equation}
We therefore conclude that 
\begin{equation}
\Big|\big<0_{\rm out}\big|0_{\rm in}\big>\Big|^2 <1\; .
\label{eq:LI-1}
\end{equation}
In other words, the probability that the vacuum stays empty is
strictly smaller than unity. Following the conventions
of~\cite{ItzykZ1}, we can write the vacuum-to-vacuum transition
amplitude as
\begin{equation}
\big<0_{\rm out}\big|0_{\rm in}\big>\equiv e^{i{\cal V}[\rho]} \; ,
\label{eq:LI-2}
\end{equation}
where $i{\cal V}[\rho]$ compactly represents the sum of the connected
vacuum-vacuum diagrams in the presence of the external (in our case,
strong, time dependent, colored) source $\rho$. Therefore, the
inequality (\ref{eq:LI-1}) means that vacuum-vacuum diagrams have a
non-zero imaginary part, since $\big|\big<0_{\rm out}\big|0_{\rm
in}\big>\big|^2 = \exp(-2\,{\rm Im}\,{\cal V}[\rho])$. In stark
contrast, for a field theory without external sources,
eq.~(\ref{eq:LI-1}) would be an equality, and the vacuum-vacuum
diagrams would be purely real, thereby only giving a pure phase for
the vacuum-to-vacuum amplitude in eq.~(\ref{eq:LI-2}).  They therefore
do not contribute to the probabilities for producing particles.

Eq.~(\ref{eq:LI-1}) tells us that one has to be more careful in field theories with external sources. To
illustrate how particle production works in such theories, we shall,
for simplicity, consider a real scalar field with cubic
self-interactions, coupled to an external source. (The lessons we draw
carry over straightforwardly to QCD albeit their implementation is in
practice significantly more complex.)  The Lagrangian density is
\begin{equation}
{\cal L}\equiv\frac{1}{2}\partial_\mu\phi\,\partial^\mu\phi 
-\frac{1}{2}m^2\phi^2-\frac{g}{3!}\phi^3 +\rho\,\phi\; .
\label{eq:LI-3}
\end{equation}
Note that the coupling $g$ in this theory has dimensions of the mass;
and that the theory is super-renormalizable in $n= 4$ dimensions.  The
source densities $\rho(x)$ can be envisioned as the scalar analogue of
the sum of two source terms $\rho(x)=\rho_1(x)+\rho_2(x)$
corresponding respectively in the CGC framework to the recoil-less
color currents of the two hadronic projectiles.

Let us now consider how the perturbative expansion for such a theory
looks like in weak coupling. The power of a generic simply connected
diagram is given simply by
\begin{equation}
g^{n_{_E}+2(n_{_L}-1)}\big(g\,\rho\big)^{n_{\rho}}\,,
\label{eq:LI-3.1}
\end{equation}
where $n_{_E}$ are the number of external lines, $n_{_L}$ the number
of loops and $n_{\rho}$ the number of sources. For vacuum--vacuum
graphs, $n_{_E}=0$. As $\rho\sim 1/g$, the power counting for a theory
with strong sources is given entirely by an expansion in the number of
loops. In particular, at tree level ($n_{_L}=0$), the vacuum--vacuum
graphs are all of order $1/g^2$.  The tree graphs contributing to the
connected vacuum-vacuum amplitude in eq.~(\ref{eq:LI-1}) can be
represented as \setbox1=\hbox to
7cm{\resizebox*{7cm}{!}{\includegraphics{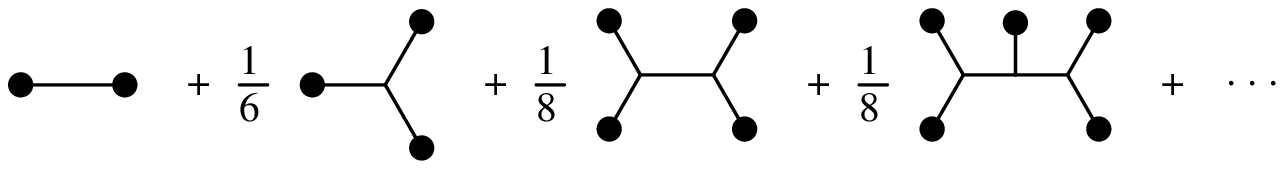}}}
\begin{equation}
i{\cal V}[j]\equiv i\sum_{\rm conn}V = \quad\;\raise -3.5mm\box1
\label{eq:LI-4}
\end{equation}
There are also loop contributions in this expression which we have not
represented here.

To proceed with the perturbative computation, we need to consider the
analog of the well known Cutkosky rules for this case.  For each
diagram in the computation, begin by assigning for each vertex and
source, two kinds of vertices denoted by $+$ or $-$. A vertex of type
$+$ is the ordinary vertex and appears with a factor $-ig$ in Feynman
diagrams. A vertex of type $-$ is the opposite\footnote{Because the
coupling constant $g$ is real in an unitary theory, the vertex of type
$-$ is also the complex conjugate of the vertex of type $+$.} of a $+$
vertex, and its Feynman rule is $+ig$. Likewise, for insertions of the
source $\rho$, insertions of type $+$ appear with the factor
$+i\rho(x)$ while insertions of type $-$ appear instead with
$-i\rho(x)$.  Thus for each Feynman diagram $iV$ in
eq.~(\ref{eq:LI-3}), containing only $+$ vertices and sources (denoted
henceforth as $iV_{\{+\cdots+\}}$) contributing to the sum of
connected vacuum-vacuum diagrams, we obtain a corresponding set of
diagrams $iV_{\{\epsilon_i\}}$ by assigning the symbol $\epsilon_i =
\pm $ to the vertex $i$ of the original diagram (and connecting a
vertex of type $\epsilon$ to a vertex of type $\epsilon^\prime$ with
a propagator $G^0_{\epsilon\epsilon^\prime}$ -- to be discussed further
shortly).

The generalized set of diagrams therefore includes $2^n$ such diagrams
if the original diagram had $n$ vertices and sources. Using
recursively the so-called ``largest time
equation''~\cite{t'HooV1,Veltm1}, one obtains the identity,
\begin{equation}
iV_{\{+\cdots+\}}+iV_{\{-\cdots-\}}
=-2\,{\rm Im}\,V
= -\sum_{\{\epsilon_i\}^\prime}
iV_{\{\epsilon_i\}}\; ,
\label{eq:LI-5}
\end{equation}
where the prime in the sum means that we sum over all the combinations
of $\epsilon_i$'s, except the two terms where the vertices and sources
are all of type $+$ or all of type $-$. (There are therefore $2^n-2$
terms in this sum.)

We now need to specify the propagators connecting the $\pm$ vertices
and sources.  The usual Feynman (time-ordered) free propagator is the
propagator connecting two vertices of type $+$, i.e. $G^0_{++}$. It
can be decomposed as
\begin{equation} 
G^0_{++}(x,y)
\equiv \theta(x^0-y^0)\,G^0_{-+}(x,y)+\theta(y^0-x^0)\,G^0_{+-}(x,y)\,,
\end{equation}
which defines the propagators $G^0_{-+}$ and $G^0_{+-}$. Likewise, the
anti--time-ordered free propagator $G^0_{--}$ is defined
as~\footnote{The notations for the propagators are those that appear
in the Schwinger-Keldysh formalism developed initially for field
theories at finite temperature (see \cite{Schwi1,Keldy1}). This
identification, as we shall see later, is not accidental.  The
propagators so defined are not independent; they are related by the
identity $G^0_{++}+G^0_{--}=G^0_{-+}+G^0_{+-}$.}
\begin{equation}
G^0_{--}(x,y)
\equiv \theta(x^0-y^0)\,G^0_{+-}(x,y)+\theta(y^0-x^0)\,G^0_{-+}(x,y)\; .
\end{equation}

The Fourier transforms of the free propagators
$G^0_{\epsilon\epsilon^\prime}$ for our scalar theory are
\begin{eqnarray}
&&
G^0_{++}(p)=\frac{i}{p^2-m^2+i\epsilon}\;,\quad
G^0_{--}(p)=\frac{-i}{p^2-m^2-i\epsilon}\; ,
\nonumber\\
&&
G^0_{-+}(p)=2\pi\theta(p^0)\delta(p^2-m^2)\; ,\;
G^0_{+-}(p)=2\pi\theta(-p^0)\delta(p^2-m^2)\; .
\label{eq:LI-6}
\end{eqnarray}

For a given term in the right hand side of eq.~(\ref{eq:LI-5}), one
can divide the diagram in several regions, each containing only $+$ or
only $-$ vertices and sources. (There is at least one external source
in each of these regions because of energy conservation constraints.)
The $+$ regions and $-$ regions of the diagram are separated by a
``cut'', and one thus obtains a ``cut vacuum-vacuum diagram".  At tree
level, the first terms generated by these cutting rules (applied to
compute the imaginary part of the sum of connected vacuum-vacuum
diagrams in eq.~(\ref{eq:LI-5})) are \setbox1=\hbox to
8cm{\resizebox*{8cm}{!}{\includegraphics{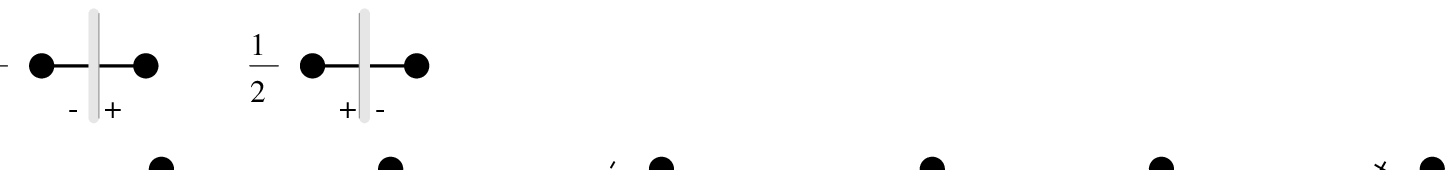}}}
\begin{eqnarray}
2\,{\rm Im}\,\sum_{\rm conn}V&=&\quad\;\raise -24mm\box1\nonumber\\
&&\qquad\qquad\qquad\;\, + \cdots \nonumber \\
&\equiv& \frac{1}{g^2}\sum_r b_r\; .
\label{eq:LI-7}
\end{eqnarray}
The $+$ and $-$ signs adjacent to the grey line in each diagram here
indicate the side on which the set of $+$ and $-$ vertices is located.
As one can see, there are cuts intercepting more than one
propagator. The sum of the diagrams with $r$ cut propagators is
denoted by $b_r/g^2$ -- the identity in eq.~(\ref{eq:LI-7}) (and
eq.~(\ref{eq:LI-5})) is a statement of unitarity. These $b_r$ are
sometimes called ``combinants" in the
literature~\cite{GyulassyKauffman}.

It is important to note that cut connected vacuum-vacuum diagrams
would be zero in the vacuum because energy cannot flow from one side
of the cut to the other in the absence of external sources. This is of
course consistent with a pure phase in eq.~(\ref{eq:LI-5}). This
constraint on the energy flow is removed if the fields are coupled to
{\bf time-dependent} external sources. Cut vacuum-vacuum diagrams, and
therefore the imaginary part of vacuum-vacuum diagrams, differ from
zero in this case.

\begin{figure}[htbp]
\setbox1=\hbox to 8cm{\resizebox*{8cm}{!}{\includegraphics{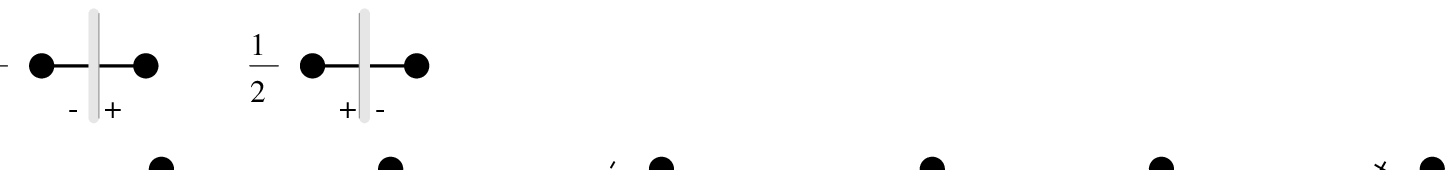}}}
\setbox2=\hbox to 8cm{\resizebox*{8cm}{!}{\includegraphics{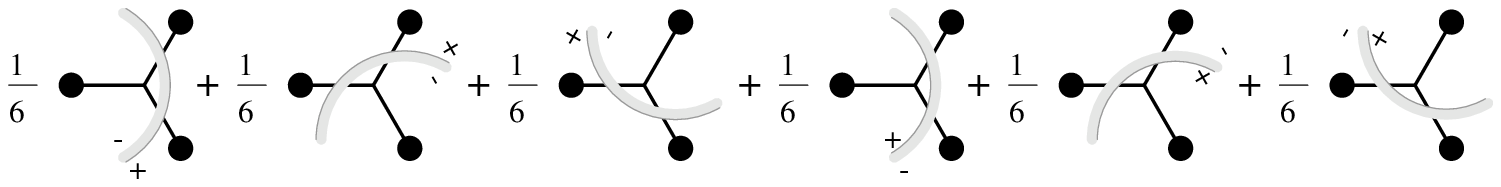}}}
\setbox3=\hbox to 5.3cm{\resizebox*{5.3cm}{!}{\includegraphics{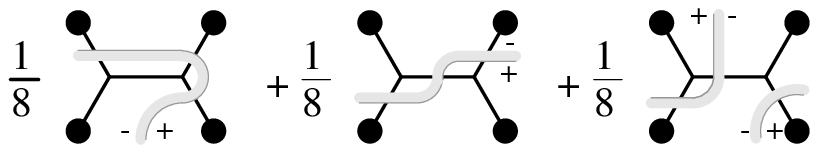}}}
\begin{eqnarray}
\frac{b_1}{g^2}&=&\quad\;\raise -13mm\box1 \nonumber \\
&+&\cdots\nonumber\\
&&\nonumber\\
\frac{b_2}{g^2}&=&\quad\;\raise -4.5mm\box2 \nonumber \\
&+&\cdots\nonumber\\
&&\nonumber\\
\frac{b_3}{g^2}&=&\quad\;\raise -3.5mm\box3 \nonumber \\
&+&\cdots\nonumber
\end{eqnarray}
\caption{\label{fig:b123} Examples of cut diagrams contributing to $b_1$, $b_2$
and $b_3$.}
\end{figure}

We now turn to the probabilities for producing $n$ particles. The
probability to produce one particle from the vacuum can be
parameterized as
\begin{equation}
P_1 = e^{-\frac{1}{g^2}\sum_r b_r} \; \frac{b_1}{g^2}\; ,
\label{eq:LI-8.1}
\end{equation}
where $b_1$, a series in $g^{2n}$ ($n\geq 0$) is obtained by summing
the 1-particle cuts through connected vacuum-vacuum diagrams.  The
exponential prefactor is the square of the sum of all the
vacuum-vacuum diagrams, which arises in any transition
probability. The probability $P_2$ for producing two particles from
the vacuum contains two pieces. One is obtained by squaring the
$b_1/g^2$ piece of the probability for producing one particle
(dividing by $2\!$ for identical particles) -- in this case, the two
particles are produced independently from one another.  The other
$b_2/g^2$ is a ``correlated" contribution from a 2-particle cut
through connected vacuum-vacuum diagrams. We therefore obtain
\begin{equation}
P_2=  e^{-\frac{1}{g^2}\sum_r b_r} \; \left[\frac{1}{2!}\frac{b_1^2}{g^4} 
+ \frac{b_2}{g^2}\right]\; .
\label{eq:LI-8.2}
\end{equation}
In a similar vein, the probability $P_3$ can be shown to consist of
three pieces. One (``uncorrelated") term is the cube of $b_1/g^2$
(preceded by a symmetry factor $1/3!$). Another is the combination
$b_1 b_2 / g^4$, corresponding to the production of two particles in
the same subdiagram with the third produced independently from the
first two. Finally, there is the ``correlated" three particle
production probability $b_3/g^2$ corresponding to the production of
three particles from the same diagram. The sum of these three pieces
is thus
\begin{equation}
P_3 = e^{-\frac{1}{g^2}\sum_r b_r}  \; \left[\frac{1}{3!}\frac{b_1^3}{g^6} 
+\frac{b_1b_2}{g^4}+ \frac{b_3}{g^2}\right]\; .
\label{eq:LI-8.3}
\end{equation}
Some of the graphs contributing to $b_1$, $b_2$ and $b_3$ are shown in
fig.~\ref{fig:b123}.
Following this line of inductive reasoning, one obtains a general
formula for the production of $n$ particles
\begin{equation}
P_n=e^{-\frac{1}{g^2}\sum_r b_r}  \; \sum_{p=0}^n \frac{1}{p!}
\sum _{\alpha_1+\cdots+\alpha_p=n}
\frac{b_{\alpha_1}\cdots b_{\alpha_p}}{g^{2p}}\; .
\label{eq:LI-8.f}
\end{equation}
for any $n$. In this formula, $p$ is the number of disconnected
subdiagrams producing the $n$ particles, and $b_r/g^2$ denotes the sum
of all $r$-particle cuts through the connected vacuum-vacuum diagrams.
This formula gives the probability of producing $n$ particles to all
orders in the coupling $g$ in a field theory with strong external
sources. It is important to realize that all the details of the
dynamics of the theory under consideration are hidden in the numbers
$b_r$, and that eq.~(\ref{eq:LI-8.f}) is a generic form for transition
probabilities when many disconnected graphs as well as vacuum-vacuum
graphs contribute. This formula is therefore equally valid for QCD.

It is useful to introduce a generating function for these
probabilities,
\begin{equation}
F(z) \equiv
 \sum_{n=1}^\infty z^n\, P_n =
\exp\Big[ \frac{1}{g^2} \sum_r b_r \left(z^r-1\right)\Big]\, .
\label{eq:LI-8.g}
\end{equation}
One can use this object in order to compute moments of the
distribution of probabilities
\begin{eqnarray}
&&\langle n\rangle = (\ln F)^\prime(z=1) = \frac{1}{g^2} \sum_r r\, b_r\,.\nonumber\\
&&\langle n^2\rangle -\langle n\rangle^2 = (\ln F)^{\prime\prime}(z=1) = \frac{1}{g^2}\sum_r r^2\, b_r \;,
\end{eqnarray}
where each ``prime" denotes a derivative with respect to $z$. Note
that $F(z=1) = \sum_n P_n = 1$. This demonstrates explicitly that the
exponential prefactor in eq.~(\ref{eq:LI-8.f}) is essential for the
theory to be unitary. Though we derived eqs.~(\ref{eq:LI-8.f}) and
~(\ref{eq:LI-8.g}) independently, we were alerted by
Dremin~\cite{Dremin} that an earlier version of the formulas in
eqs.~(\ref{eq:LI-8.f}) and ~(\ref{eq:LI-8.g}) was derived by Gyulassy
and Kauffmann \cite{GyulassyKauffman} nearly 30 years ago also using general
combinatoric arguments that did not rely on specific dynamical
assumptions.

These combinatoric rules for computing probabilities (and moments
thereof) in field theory with strong external sources can be mapped on
to the AGK cutting rules derived in the context of reggeon field
theory~\cite{AbramGK1} by writing eq.~(\ref{eq:LI-8.f}) as
\begin{equation}
P_n=\sum_{p=0}^n {P}^{(c)}_{n,p}\; ,
\end{equation}
where ${P}^{(c)}_{n,p}$ denotes the probability of producing $n$
particles in $p$ cut sub-diagrams. One can ask directly what the
probability of $p$ cut sub-diagrams is by summing over $n$ to obtain
\begin{equation}
{\cal R}_p\equiv\sum_{n=p}^{+\infty}{P}^{(c)}_{n,p}=
\frac{1}{p!}\left(\frac{\sum_r b_r}{g^2}\right)^p\;e^{-\frac{1}{g^2}\sum_r b_r}\; .
\label{eq:LI-9}
\end{equation}
This is a Poisson distribution, which is unsurprising in our
framework, because disconnected vacuum-vacuum graphs are uncorrelated
by definition. The average number of such cut diagrams is simply
\begin{equation}
\big<n_{\rm cut}\big>\equiv\sum_{p=0}^{+\infty}p{\cal R}_p=
\frac{1}{g^2}\;\sum_r b_r\; .
\label{eq:LI-10}
\end{equation}
An exact identification with Ref.~\cite{AbramGK1} is obtained by
expanding the exponential in eq.~(\ref{eq:LI-9}) to order $m-p$, and
defining
\begin{equation}
{\cal R}_{p,m}=\frac{1}{(m-p)!}\frac{1}{p!}\;
\left(-\frac{\sum_r b_r}{g^2}\right)^{m-p}\left(\frac{\sum_r b_r}{g^2}\right)^p\; ,
\end{equation}
where ${\cal R}_{p,m}$ is the probability of having $p$ cut
sub-diagrams out of $m$ sub-diagrams (with $m-p$ being the number of
uncut diagrams). This distribution of probabilities can be checked to
satisfy the relations
\begin{eqnarray}
&&
\mbox{if\ }m\ge2\;,\qquad \sum_{p=1}^m p {\cal R}_{p,m} =0\; ,
\nonumber\\
&&
\mbox{if\ }m\ge3\;,\qquad \sum_{p=2}^m p(p-1) {\cal R}_{p,m} =0\; ,
\nonumber\\
&&
\cdots
\label{eq:LI-11}
\end{eqnarray}
This set of identities is strictly equivalent to the eqs.~(24) of
Ref.~\cite{AbramGK1} where $m-p$ and $p$ there are identified as the
numbers of uncut and cut reggeons respectively. The first relation
means that diagrams with two or more subdiagrams cancel in the
calculation of the multiplicity. These relations are therefore a
straightforward consequence of the fact that the distribution of the
numbers of cut subdiagrams is a Poisson distribution. They do not
depend at all on whether these subdiagrams are ``reggeons'' or not. In
the AGK approach, the first identity in eq.~(\ref{eq:LI-11}) suggests
that the average number of cut reggeons $\langle n_{\rm cut}\rangle$
can be computed from diagrams with one cut reggeon alone. The average
multiplicity satisfies the relation
\begin{equation}
\langle n\rangle = \langle n_{\rm cut}\rangle \, \langle n\rangle_1 \,,
\end{equation}
where $\langle n\rangle_1$ is the average number of particles in one
cut reggeon. Computing this last quantity of course requires a
microscopic model of what a reggeon is.

Before going on, it is useful to summarize what we have learnt at this
stage about field theories with strong external sources. We derived a
general formula in eq.~(\ref{eq:LI-8.f}) for the probability to
produce $n$ particles in terms of cut connected vacuum-vacuum
diagrams, where $b_r$ is the sum of the terms with $r$ cuts. This
formula is a purely combinatoric expression; it does not rely on the
microscopic dynamics generating the $b_r$. Nevertheless, it tells us
several things that were not obvious. Firstly, the probability
distribution in eq.~(\ref{eq:LI-8.f}) is not a Poisson distribution,
{\it even at tree level}, if any $b_r \neq 0$ for $r >1$. It is often
assumed that classical dynamics is Poissonian. We see here that the
non-trivial correlations (symbolized by non-zero $b_r$ terms with $r
>1$ ) in theories with self-interacting fields can produce significant
modifications of the Poisson distribution.

Another immediate observation is that even the probability to produce
one particle (eq.~(\ref{eq:LI-8.1})) is completely non-perturbative in
the coupling constant $g$ for arbitrarily small coupling. In other
words, $P_1$ cannot be expressed as an analytic expansion in powers of
$g$.  Therefore, while weak coupling techniques are certainly valid,
such theories (the CGC for instance) are always
non-perturbative. Interestingly, we will see shortly that a simple
expansion in powers of the coupling exists for moments of the
probability distribution. Finally, we saw that there was a simple
mapping between the cutting rules first discussed in
Ref.~\cite{AbramGK1} and those for cut connected vacuum-vacuum graphs
in field theories with strong sources.

In the rest of this lecture, we shall sketch the derivation of
explicit expressions for the $n$-particle probabilities and for the
first moment of the multiplicity distribution. Specifically, we will
outline an algorithm to compute the average multiplicity up to
next-to-leading order in the coupling constant.  The probability for
producing $n$ particles is given by the expression
\begin{equation}
P_n = \frac{1}{n!}
\int \left[\prod_{i=1}^n \frac{d^3 \p_i}{(2\pi)^3 2E_i}\right] 
\left|\big<\p_1\cdots \p_n{}_{\rm out}\big|0_{\rm in}\big>\right|^2 \; ,
\label{eq:LI-12}
\end{equation}
where $E_i \equiv \sqrt{{{\p_i}}^2 + m^2}$. The well known
Lehman--Symanzik--Zimmerman (LSZ) reduction formula~\cite{ItzykZ1}
relates the transition amplitude for producing $n$ particles from the
vacuum to the residue of the multiple poles of Green's functions of
the interacting theory. It can be expressed as
\begin{equation}
\big<\p_1\cdots \p_n{}_{\rm out}\big|0_{\rm in}\big>
=\frac{1}{Z^{n/2}}
\int 
\left[\prod_{i=1}^n d^4x_i \; e^{ip_i\cdot x_i}\;(\square_{x_i}+m^2)
\frac{\delta}{i\delta \rho(x_i)}\right]
\;
e^{i{\cal V}[\rho]}\; ,
\label{eq:LI-13}
\end{equation}
where the factors of $Z$ correspond to self-energy
corrections\footnote{More precisely, they are the wavefunction
renormalization factors.} on the cut propagators of the vacuum-vacuum
diagrams. Substituting the r.h.s. of this equation into
eq.~(\ref{eq:LI-12}), and noting that
\begin{equation}
G_{+-}^0(x,y) = 
\int \frac{d^3 \p_i}{(2\pi)^3 \,2E_i} e^{ip\cdot (x-y)} \; ,
\label{eq:pm-propagator}
\end{equation}
is the Fourier transform of the propagator given in
eq.~(\ref{eq:LI-6}), we can write the probability $P_n$ directly as
\begin{eqnarray}
P_n=
\frac{1}{n!}
{\cal D}^n[\rho_+,\rho_-]
\;
\left.
e^{i{\cal V}[\rho_+]}\;e^{-i{\cal V}^*[\rho_-]}
\right|_{\rho_+=\rho_-=\rho}\; ,
\label{eq:LI-14}
\end{eqnarray}
where ${\cal D}[\rho_+,\rho_-]$ is the operator
\begin{equation}
{\cal D}\equiv
\int d^4x\, d^4y \;
ZG_{+-}^0(x,y) \; \frac{\square_x+m^2}{Z}\;\frac{\square_y+m^2}{Z}
\frac{\delta}{\delta \rho_+(x)}\frac{\delta}{\delta \rho_-(y)}\; .
\label{eq:LI-15}
\end{equation}
The sources in the amplitude and the complex conjugate amplitude are
labeled as $\rho_+$ and $\rho_-$ respectively to ensure that the
functional derivatives act only on one of the two factors.

A useful and interesting identity is
\begin{equation}
e^{{\cal D}[\rho_+,\rho_-]} \;
e^{i{\cal V}[\rho_+]}\;e^{-i{\cal V}^*[\rho_-]}
=
e^{i{\cal V}_{_{SK}}[\rho_+,\rho_-]}\; ,
\label{eq:LI-16}
\end{equation}
where $i{\cal V}_{_{SK}}[\rho_+,\rho_-]$ is the sum of all connected
vacuum-vacuum diagrams in the Schwinger-Keldysh formalism
\cite{Schwi1,Keldy1}, with the source $\rho_+$ on the upper branch of
the contour and likewise, $\rho_-$ on the lower branch.  When
$\rho_+=\rho_-=\rho$, it is well known that this sum of all such
connected vacuum-vacuum diagrams is zero. The generating function
$F(z)$, from eqs.~(\ref{eq:LI-8.g}) and (\ref{eq:LI-14}) is simply
\begin{equation}
F(z) = e^{zD}\;\left.
e^{i{\cal V}[\rho_+]}\;e^{-i{\cal V}^*[\rho_-]}
\right|_{\rho_+=\rho_-=\rho}\; .
\label{eq:LI-17}
\end{equation}
From the expression for the operator $D$ in eq.~(\ref{eq:LI-15}), it
is clear that $F(z)$ can be formally obtained by substituting the
off-diagonal propagators $G_{\mp,\pm}^0\rightarrow z\, G_{\mp,\pm}^0$
in the usual cut vacuum-vacuum diagrams.

We shall now proceed to discuss how one computes the average
multiplicity ($F^\prime(z=1)$) of produced particles.  From
eqs.~(\ref{eq:LI-17}) and (\ref{eq:LI-15}), we obtain
\begin{equation}
\big<n\big>
=
\int d^4x\, d^4y\;
ZG_{+-}^0(x,y)
\;\left[
\Gamma^{(+)}(x)\Gamma^{(-)}(y)+\Gamma^{(+-)}(x,y)
\right]_{\rho_+=\rho_-=\rho}\; ,
\label{eq:LI-18}
\end{equation}
where $\Gamma^{(\pm)}$ and $\Gamma^{(+-)}$ are the 1- and 2-point
amputated Green's functions in the Schwinger-Keldysh formalism:
\begin{eqnarray}
&&
\Gamma^{(\pm)}(x)\equiv \frac{\square_x+m^2}{Z}\;
\frac{\delta i{\cal V}_{_{SK}}[\rho_+,\rho_-]}{\delta \rho_\pm(x)}\; ,
\nonumber\\
&&
\Gamma^{(+-)}(x,y)
\equiv
\frac{\square_x+m^2}{Z}\;\frac{\square_y+m^2}{Z}\;
\frac{\delta^2 i{\cal V}_{_{SK}}[\rho_+,\rho_-]}{\delta \rho_+(x)\delta \rho_-(y)}\; .
\label{eq:LI-19}
\end{eqnarray}
Diagrammatically, $\big<n\big>$ can be represented as
\setbox1\hbox to 4cm{\hfil\resizebox*{4cm}{!}{\includegraphics{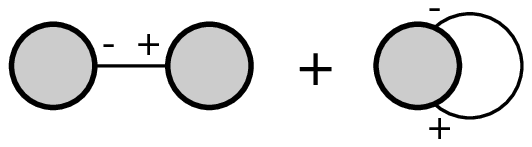}}}
\begin{equation}
\big<n\big>=\quad\raise -5mm\box1\quad .
\label{eq:LI-20}
\end{equation}
Unlike the probabilities, there is a well defined power counting for
the moments of the multiplicity distribution. This is simply because
the overall ``absorption factor" $\exp(-{\sum_r b_r}/{g^2})$ present
in the computation each probability, cancels when one computes
averaged quantities. This is a crucial simplification, because it
means that the moments of the distribution have a
sensible\footnote{The usual caveats, about the convergence of such
series and issues related to their Borel summability, apply here as
well.} perturbative expansion as a series in powers of $g^2$.

At leading order in the coupling constant, ${\cal O}(g^{-2}
(g\rho)^n)$, only the left diagram in eq.~(\ref{eq:LI-20})
contributes. The right diagram, that contains the connected 2-point
function, is a one loop diagram; in our power counting (see
eq.~(\ref{eq:LI-3.1})) it starts at order ${\cal O}(g^0 (g\rho)^n)$.
The lowest order, where we need only tree level diagrams, can
therefore be expressed as \setbox1=\hbox to
1.8cm{\hfil\resizebox*{1.6cm}{!}{\includegraphics{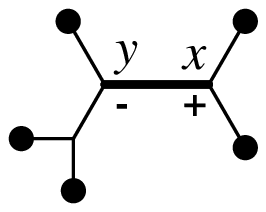}}}
\begin{equation}
\big<n\big>_{_{LO}}
=\sum_{+/-} \quad\raise -7mm\box1\quad,
\label{eq:LI-21}
\end{equation}
where the sum is over all the tree diagrams on the left and on the
right of the propagator $G_{+-}^0$ (represented in boldface) as well
as a sum over the labels $+/-$ of the vertices whose type is not
written explicitly. At this order, the mass in $G_{+-}^0$ is simply
the bare mass, and $Z=1$.

The diagrams in eq.~(\ref{eq:LI-21}) can be computed using the
Cutkosky rules we discussed previously. Beginning with one of the
``leaves'' of the tree (attached to the rest of the diagram by a $+$
vertex for instance), one has two contributions $G_{++}^0$ and
$-G_{+-}^0$ for the propagators connecting it to the vertex just below.
(The source can be factored out, because we set
$\rho_+=\rho_-=\rho$.) This difference in the propagators gives
\begin{equation}
G^0_{++}-G^0_{+-}=G^0_{_R}\; ,
\label{eq:LI-22}
\end{equation}
where $G^0_{_R}$ is the free {\sl retarded propagator}. (Likewise,
$G^0_{-+}-G^0_{--}=G^0_{_R}$.) Repeating this procedure recursively,
propagators from all the ``leaves" down to the root are converted into
retarded propagators. It is well known that the retarded solution
$\phi_c(x)$ of the classical equations of motion with the initial
conditions $\lim_{x^0\rightarrow -\infty}\phi_c(x)=0$ and
$\lim_{x^0\rightarrow -\infty}\partial^0 \phi_c(x)=0$ can be expressed
as a sum of tree diagrams built with retarded propagators. The sum
over all the trees on each side of the cut in eq.~(\ref{eq:LI-21}) can
therefore be identified as \setbox1=\hbox to
1.55cm{\hfil\resizebox*{1.55cm}{!}{\includegraphics{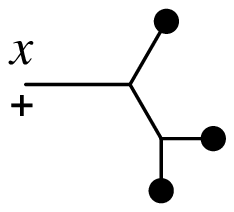}}}
\setbox2=\hbox to
1.55cm{\hfil\resizebox*{1.55cm}{!}{\includegraphics{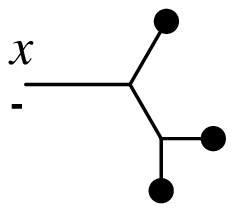}}}
\begin{equation}
\phi_c(x)=\sum_{+/-}\;\;\raise -8mm\box1 \quad
= \sum_{+/-}\;\;\raise -8mm\box2 \,.
\label{eq:LI-23}
\end{equation}

From this discussion and eq.~(\ref{eq:LI-18}), the leading order
inclusive multiplicity can be expressed as
\begin{equation}
\big<n\big>_{_{LO}}=\int\frac{d^3\p}{(2\pi)^32E_p}\;
\left|\int d^4x \;e^{ip\cdot x}\;(\square+m^2)\phi_c(x)\right|^2\; .
\label{eq:LI-24}
\end{equation}
Using the identity $e^{ip\cdot x} (\partial_0^2+E_p^2)\phi_c(x) =
\partial_0 \left(e^{ip\cdot
x}\big[\partial_0-iE_p\big]\phi_c(x)\right)$ and the boundary
conditions obeyed by the retarded classical field $\phi_c(x)$, one
obtains
\begin{equation}
E_p\frac{d\langle n\rangle_{\rm LO}}{d^3 p}=\frac{1}{16\pi^3}
\left|
\lim_{x^0\rightarrow +\infty}\int d^3 x\, e^{ip\cdot x}\left[\partial_{x^0} - iE_p\right]\phi_c(x)\right|^2\; .
\label{eq:LI-25}
\end{equation}
The corresponding formula for gluon production in heavy ion collisions
in the Color Glass Condensate framework is
\begin{eqnarray}
E_p\frac{d\langle n\rangle_{\rm LO}}{d^3 p}
&=&
\frac{1}{16\pi^3} 
\lim_{x^0,y^0\rightarrow +\infty}
\int d^3 x\, d^3 y \,e^{ip\cdot (x-y)} 
(\partial_{x^0} - iE_p)(\partial_{y^0} +iE_p)\nonumber \\
&\times&
\sum_{{\rm phys.}\,\lambda}\epsilon_\mu^\lambda(\p)\, 
{\epsilon^\star}_\nu^\lambda(\p)\;  A_c^\mu (x) A_c^\nu (y) \; ,
\label{eq:LI-26}
\end{eqnarray}
where $\epsilon_\mu^\lambda$ is the polarization vector for the
produced gluon.  This is precisely the expression that was computed in
previous real time numerical simulations of Yang--Mills equations {\it
for each configuration of color sources in each of the nuclei}. To
compute the distribution of gluons, we need to average over the
distribution over all possible color sources as specified in
eq.~(\ref{eq:LI-0}). We will discuss results from these simulations
further in Lecture II.

The leading order result in eqs.~(\ref{eq:LI-25}) and (\ref{eq:LI-26})
is well known. We shall now discuss the computation to next-to-leading
order in the coupling -- to order ${\cal O}(g^0 (gj)^n)$. At this
order, both terms in eq.~(\ref{eq:LI-18}) contribute to the
multiplicity. The right diagram in eq.~(\ref{eq:LI-20}) contributes
with the blob evaluated at tree level, \setbox1=\hbox to
2cm{\hfil\resizebox*{2cm}{!}{\includegraphics{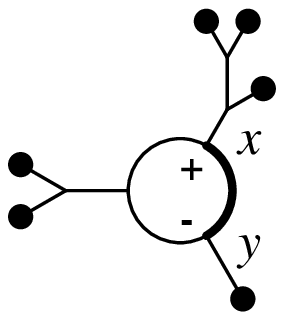}}}
\begin{equation}
\raise -7.5mm\box1 \, .
\label{eq:LI-27}
\end{equation}
This contribution to the inclusive multiplicity is analogous to that
of quark-anti-quark pair production or gluon pair production to the
respective average multiplicities for these quantities.  The left
diagram in eq.~(\ref{eq:LI-20}), at this order, contains 1-loop
corrections to diagrams of the kind displayed in
eq.~(\ref{eq:LI-21}). A blob on one side of the cut in
eq.~(\ref{eq:LI-20}) is evaluated at the 1 loop level (corresponding
to the contribution from one loop correction to the classical field)
while the other blob is evaluated at tree level (corresponding to the
contribution from the classical field itself). This can be represented
as \setbox1=\hbox to
1.8cm{\hfil\resizebox*{1.8cm}{!}{\includegraphics{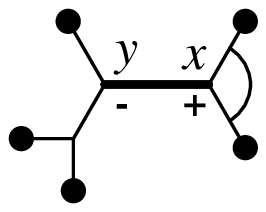}}}
\begin{equation}
\raise -8mm\box1
\label{eq:LI-28}
\end{equation}
The inclusive multiplicity at NLO includes contributions from both
eqs.~(\ref{eq:LI-27}) and (\ref{eq:LI-28}).

To evaluate the diagram in eq.~(\ref{eq:LI-27}), one needs to compute
the propagator $G_{+-}(x,y)$ in the presence of the background field
$\phi_c$.  This can be done by solving a Lippmann-Schwinger equation
for $G_{+-}$~\cite{GelisKL1,GelisV1,BaltzGMP1}. In practice, numerical
solutions of this equation can be obtained only for {\sl retarded} or
{\sl advanced} Green's functions in the background field.  It turns
out that one can express $G_{+-}$ in terms of these as
\begin{equation}
G_{+-}=G_{_R} G_{_R}^0{}^{-1} G_{+-}^0 G_{_A}^0{}^{-1}G_{_A} \, ,
\label{eq:LI-29}
\end{equation}
where 
\begin{eqnarray}
&&
G_{_R}\equiv G_{_R}^0 + G_{_R}^0 T_{_R} G_{_R}^0\; ,
\nonumber\\
&&
G_{_A}\equiv G_{_A}^0 + G_{_A}^0 T_{_A} G_{_A}^0\; .
\label{eq:LI-30}
\end{eqnarray}
Here $G_{_R}^0$ ($G_{_A}^0$) is the free retarded (advanced)
propagator and $T_{_R}$ ($T_{_A}$) is the retarded (advanced)
scattering $T$-matrix.  Substituting eq.~(\ref{eq:LI-30}) in
eq.~(\ref{eq:LI-29}) and using the resulting expression in the second
term of eq.~(\ref{eq:LI-18}), the contribution of this term to the NLO
multiplicity can be expressed as
\begin{equation}
\big<n\big>_{_{NLO}}^{(1)}
=\int\frac{d^3\p}{(2\pi)^3 2E_p}
\int \frac{d^3\q}{(2\pi)^3 2E_q}\;\left|T_{_R}(p,-q)\right|^2\; .
\label{eq:LI-31}
\end{equation}
One can then show that~\cite{GelisV1}
\begin{equation}
T_{_R}(p,-q)=\lim_{x_0\to +\infty}
\int d^3\x \; e^{ip\cdot x}
\left[\partial_{x_0}-iE_p\right]
\eta_q(x)\; ,
\label{eq:LI-32}
\end{equation}
where $\eta_q(x)$ is a small fluctuation field about $\phi_c(x)$ and
is the {\it retarded} solution of the partial differential equation
\begin{equation}
\big(\square+m^2+g\phi_c(x)\big)\eta_q(x)=0\; ,
\label{eq:LI-33}
\end{equation}
with the initial condition $\eta_q(x)=e^{iq\cdot x}$ when $x_0\to
-\infty$. Note here that $g$ has the dimension of a mass. Note also
that, despite being similar, the equation for $\eta$ is not the
classical equation of motion but is instead the equation of motion of
a small fluctuation.  This NLO contribution to the inclusive
multiplicity can be computed by solving an initial value problem with
boundary conditions set at $x^0\rightarrow -\infty$.

The other contribution of order ${\cal O}(g^0 (gj)^n)$ to the average
multiplicity is from the diagram in eq.~(\ref{eq:LI-28}). This
contribution can be written as
\begin{eqnarray}
\big<n\big>_{_{NLO}}^{(2)}
&=&
\displaystyle{\int \frac{d^3\p}{(2\pi)^3 2E_p}}
\Big[
\lim_{x_0\to+\infty}\int d^3\x \; e^{ip\cdot x}
\big[\partial_0-iE_p\big]\phi_c(x)
\Big]\nonumber\\
&&\qquad\qquad\times
\Big[
\lim_{x_0\to+\infty}\int d^3\x \; e^{ip\cdot x}
\big[\partial_0-iE_p\big]\phi_{c,1}(x)
\Big]^*+\mbox{c.c.}\nonumber\\
&&
\label{eq:LI-34}
\end{eqnarray}
The one loop contribution to the classical field
\setbox1=\hbox to
4cm{\resizebox*{4cm}{!}{\includegraphics{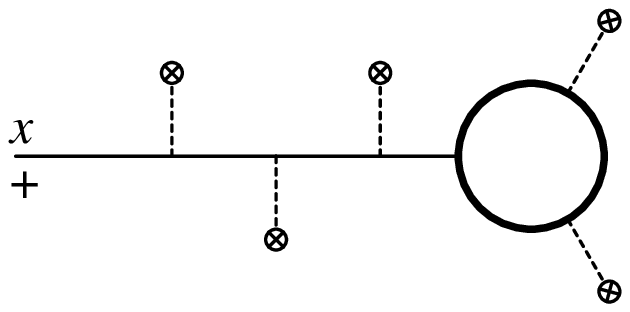}}}
\begin{equation}
\phi_{c,1}(x) = \sum_{+/-}\;\;\raise -9.5mm\box1\,,
\label{eq:LI-35}
\end{equation}
includes arbitrary insertions of the background field $\phi_c(x)$.
Following the discussion before eq.~(\ref{eq:LI-22}) of the Cutkosky
rules in this case, it can be written as~\cite{GelisV1}
\begin{equation}
\phi_{c,1}(x) =
-ig\int d^4y\;
G_{_R}(x,y)\;G_{++}(y,y)
\; .
\label{eq:LI-36}
\end{equation}
We have used here the identity $G_{++}(x,x)= G_{--}(x,x)$.  In
practice, $\phi_{c,1}(x)$ can also be obtained as the {\it retarded}
solution to the equation
\begin{equation}
(\square+m^2+g\phi_c(x))\phi_{c,1}(x)=-g\,G_{++}(x,x)\; ,
\label{eq:LI-37}
\end{equation}
with an initial condition such that $\phi_{c,1}$ and its derivatives
vanish at $x_0=-\infty$.  The source term in this equation can be
rewritten as $G_{++}(x,x)= \frac{1}{2} G_R (G_R^0)^{-1}G_S^0
(G_A^0)^{-1}G_A$, where $G_S^0(p) = 2\pi \delta(p^2-m^2)$. After a
little algebra~\cite{GelisV1}, one can show that
\begin{equation}
G_{++}(x,x)=\frac{1}{2}\int\frac{d^4q}{(2\pi)^4}\;2\pi\delta(q^2-m^2)\;
\eta_q^{(+)}(x)\eta_q^{(-)}(x)\; .
\label{eq:LI-38}
\end{equation}
Here $\eta_q^{(+)}(x)$ and $\eta_q^{(-)}(x)$ are solutions of
eq.~(\ref{eq:LI-33}) with plane wave initial conditions at
$x_0\to-\infty$ of $\eta_q^{(+)}(x)=e^{iq\cdot x}$ and
$\eta_q^{(-)}(x)=e^{-iq\cdot x}$ respectively.  We observe that
$G_{++}(x,x)$ contains ultraviolet divergences that arise from the
integration over the momentum $q$ in eq.~(\ref{eq:LI-38}). They can be
identified with the usual 1-loop ultraviolet divergences of the
$\phi^3$ field theory in the vacuum and must be subtracted
systematically in order to obtain a finite result.

To summarize, the two NLO contributions to the inclusive multiplicity,
eqs.~(\ref{eq:LI-31}) and (\ref{eq:LI-34}) can be computed
systematically as follows. One first computes the lowest order
classical field $\phi_c(x)$ by solving the classical equations of
motion, as a function of time, with the retarded boundary condition
$\phi_c(x)=0$ at $x^0=-\infty$. This computation was performed
previously in the CGC
framework~\cite{KrasnV4,KrasnV1,KrasnV2,KrasnNV1,KrasnNV2,Lappi1}. The
small fluctuation equation of motion in eq.~(\ref{eq:LI-33}) is then
solved in the background of $\phi_c(x)$, also with retarded boundary
conditions at $x^0=-\infty$ for the small fluctuation field
$\eta_q(x)$.  This is then sufficient, from eqs.~(\ref{eq:LI-32}) and
(\ref{eq:LI-31}), to compute one contribution to the NLO multiplicity.
To compute the other, solutions of the small fluctuation equations of
motion can also be used, following eq.~(\ref{eq:LI-38}), to determine
$G_{++}(x,x)$. Subsequent to this determination, the temporal
evolution of the one loop classical field can be computed by solving
eq.~(\ref{eq:LI-37}), again with retarded boundary conditions at
$x^0=-\infty$. Finally, this result can be substituted in
eq.~(\ref{eq:LI-34}) in order to compute the second contribution to
the NLO multiplicity.

Albeit involved and technically challenging, the algorithm we have
outlined is straightforward. The extension to the QCD case can be
done. Indeed, this computation is similar to a numerical computation
(performed by Gelis, Kajantie and Lappi~\cite{GelisKL1}) of the number
of produced quark pairs in the classical background field of two
nuclei. 

An interesting question we shall briefly consider now is whether we can
directly compute the generating function itself to some order in the
coupling; even a leading order computation would contain a large
amount of information. From eqs.~(\ref{eq:LI-17}) and
(\ref{eq:LI-15}), we obtain\footnote{From this relation, we see that
  the logarithm of $F(z)$ has a well defined perturbative expansion in
  powers of $g^2$ (that starts at the order $g^{-2}$), while this is
  not the case for $F(z)$ itself.}
\begin{equation}
\frac{F^\prime(z)}{F(z)}
=
\int d^4x\, d^4y \;ZG^0_{+-}(x,y)
\Big[
\Gamma^{(+)}(z|x)\Gamma^{(-)}(z|y)
+\Gamma^{(+-)}(z|x,y)
\Big]\; ,
\label{eq:LI-39}
\end{equation}
where $\Gamma^{(\pm)}(z|x)$ and $\Gamma^{(+-)}(z|x,y)$ are defined as
in eq.~(\ref{eq:LI-19}), but must be evaluated with the substitution
$G_{\mp,\pm}^0\rightarrow z\, G_{\mp,\pm}^0$ of the off-diagonal
propagators. Unsurprisingly, this equation involves the same
topologies as that for the average multiplicity in
eq.~(\ref{eq:LI-20}).  If we can compute the expression in
eq.~(\ref{eq:LI-39}) even to leading order, the generating function
can be determined directly by integration over $z$, since we know that
$F(1)=1$.

At leading order, as we have seen, only the first term in
eq.~(\ref{eq:LI-39}) contributes and (using the same trick as in
eq.~(\ref{eq:LI-25})) eq.~(\ref{eq:LI-39}) can be written as
\begin{eqnarray}
\left.
\frac{F^\prime(z)}{F(z)}
\right|_{_{\rm LO}}
\!\!\!&=&\!\!\!
\int \frac{d^3\p}{(2\pi)^3 2E_\p}\;
\Big[
\int d^3\x\; e^{ip\cdot x}\;(\partial_x^0-iE_p)\,\Phi_+(z|x)
\Big]_{x_0=-\infty}^{x_0=+\infty}
\nonumber\\
&&\qquad\qquad\qquad\times
\Big[
\int d^3\y\; e^{-ip\cdot y}\;(\partial_y^0+iE_p)\,\Phi_-(z|y)
\Big]_{y_0=-\infty}^{y_0=+\infty}\, ,
\nonumber\\
&&
\label{eq:LI-41}
\end{eqnarray}
where 
$\Phi_\pm(z|x)$ correspond to the tree diagrams 
\setbox1=\hbox to
1.55cm{\hfil\resizebox*{1.55cm}{!}{\includegraphics{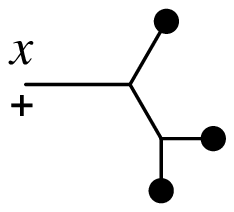}}}
\setbox2=\hbox to
1.55cm{\hfil\resizebox*{1.55cm}{!}{\includegraphics{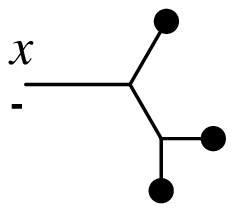}}}
\begin{equation}
\Phi_+(z|x)=\sum_{+/-}\;\;\raise -8mm\box1 \qquad
\Phi_-(z|x)= \sum_{+/-}\;\;\raise -8mm\box2
\quad,
\label{eq:LI-42}
\end{equation}
evaluated with Cutkosky's rules where the off-diagonal propagators
$G_{+-}^0$ are multiplied by a factor $z$.

We will now see why computing the generating function at leading order
is significantly more complicated than computing the average
multiplicity. The fields in eq.~(\ref{eq:LI-42}) can equivalently be
expressed as the integral equation
\begin{eqnarray}
\Phi_+(z|x)&=&i\int d^4y\; 
\Big\{G^0_{++}(x,y)\left[j(y)-\frac{g}{2}\Phi_+^2(z|y)\right]\nonumber\\
&&\qquad\qquad\qquad
-zG^0_{+-}(x,y)\left[j(y)-\frac{g}{2}\Phi_-^2(z|y)\right]
\Big\}
\nonumber\\
\Phi_-(z|x)&=&i\int d^4y\; 
\Big\{zG^0_{-+}(x,y)\left[j(y)-\frac{g}{2}\Phi_+^2(z|y)\right]\nonumber\\
&&\qquad\qquad\qquad
-G^0_{--}(x,y)\left[j(y)-\frac{g}{2}\Phi_-^2(z|y)\right]
\Big\}\; .
\label{eq:LI-43}
\end{eqnarray}
Note now that when $z=1$, $\Phi_\pm\rightarrow \phi_c$ (defined in
eq.~(\ref{eq:LI-23})) and the propagators in eqs.~(\ref{eq:LI-43}),
from eq.~(\ref{eq:LI-22}), can be rearranged to involve only the
retarded free propagator $G_R^0$. It is precisely for this reason that
the computation of the inclusive multiplicity simplifies; the field
$\phi_c$ can be determined by solving an initial value problem with
boundary conditions at $x^0\rightarrow -\infty$. This simplification
clearly does not occur when $z\neq 1$.

In order to understand the boundary conditions for $\Phi_\pm(z|x)$ in
eqs.~(\ref{eq:LI-42}) and (\ref{eq:LI-43}), we begin by expressing
them as a sum of plane waves,
\begin{eqnarray}
&&
\Phi_+(z|x)\equiv\int\frac{d^3\p}{(2\pi)^3 2E_\p}
\left\{
f_+^{(+)}(z|x^0,\p)e^{-ip\cdot x}
+
f_+^{(-)}(z|x^0,\p)e^{ip\cdot x}
\right\}\; ,
\nonumber\\
&&
\Phi_-(z|x)\equiv\int\frac{d^3\p}{(2\pi)^3 2E_\p}
\left\{
f_-^{(+)}(z|x^0,\p)e^{-ip\cdot x}
+
f_-^{(-)}(z|x^0,\p)e^{ip\cdot x}
\right\}\; .
\label{eq:LI-44}
\end{eqnarray}
Note here that $p_0\equiv\sqrt{\p^2+m^2}$ is positive. Since
$\Phi_\pm(z|x)$ does not obey the free Klein-Gordon equation, the
coefficients functions must themselves depend on time. However,
assuming that both the source $\rho(x)$ and the coupling constant $g$ are
switched off adiabatically at large negative and positive times, the
coefficient functions $f_\pm^{(\pm)}(z|x^0,\p)$ become constants in
the limit of infinite time ($x^0\to \pm \infty$). 

The technique we use for determining the boundary conditions for the
coefficients $f_\pm^{(\pm)}(z|x^0,\p)$ is reminiscent of the
derivation of Green's theorem in electrostatics. We will not go into
the derivation here (see Ref.~\cite{GelisV2} for the detailed
derivation).  The boundary conditions at $x^0=\pm \infty$ are
\begin{eqnarray}
&&
f_+^{(+)}(z|x^0=-\infty,\p)=0\; ,
\nonumber\\
&&
f_-^{(-)}(z|x^0=-\infty,\p)=0\; ,
\nonumber\\
&&
f_-^{(+)}(z|x^0=+\infty,\p)=
z\,f_+^{(+)}(z|x^0=+\infty,\p)\; ,
\nonumber\\
&&
f_+^{(-)}(z|x^0=+\infty,\p)=
z\,f_-^{(-)}(z|x^0=+\infty,\p)\; .
\label{eq:LI-45}
\end{eqnarray}
Using eqs.~(\ref{eq:LI-44}) and (\ref{eq:LI-45}), we can write
eq.~(\ref{eq:LI-41}) as
\begin{equation}
\left.
\frac{F^\prime(z)}{F(z)}
\right|_{_{LO}}
=
\int \frac{d^3\p}{(2\pi)^3 2E_\p}\;
f_+^{(+)}(z|+\infty,\p)\,f_-^{(-)}(z|+\infty,\p)\; .
\label{eq:LI-46}
\end{equation}
Therefore evaluating the generating function at leading order requires
that we know the coefficient functions at $x^0=+\infty$.

Unlike the case of partial differential equations with retarded
boundary conditions, there are no straightforward algorithms for
finding the solution with the boundary conditions listed in
eq.~(\ref{eq:LI-45}).  Methods for solving these sorts of problems are
known as ``relaxation processes''.  A fictitious ``relaxation time''
variable $\xi$ is introduced and the simulation is begun at $\xi=0$
with functions $\Phi_\pm$ that satisfy all the boundary conditions but
not the equation of motion. These fields evolve in $\xi$ with the
equation (preserving the boundary conditions for each $\xi$)
\begin{equation}
\partial_\xi \Phi_\pm = (\square_x+m^2)\Phi_\pm +
\frac{g}{2}\Phi_\pm^2-j(x)\; ,
\end{equation}
which admits solutions of the EOM as fixed points. The r.h.s. can in
principle be replaced by any function that vanishes when $\Phi_\pm$ is
a solution of the classical EOM. This function should be chosen to
ensure that the fixed point is attractive. A similar algorithm has
been developed recently to study the real time non-equilibrium
properties of quantum fields~\cite{Berges}.

Higher moments of the multiplicity distribution can also be computed
following the techniques described here. Interestingly, the variance
(at leading order) can be computed once one obtains the solutions of
the small fluctuation equations of motion.  The computation is
outlined in Ref.~\cite{GelisV2}. Thus both the leading order variance
and the NLO inclusive multiplicity can be determined simultaneously.
The variance contains useful information that can convey information
about the earliest stages of a heavy ion collision. In particular,
correlations between particles in a range of rapidity windows can
provide insight into the early stages of a heavy ion
collision~\cite{ArmestoLP}. This provides a segue for the topic of the
second lecture on the properties of the Glasma.

\section*{Lecture II: What does the Glasma look like and how does it thermalize to form a Quark Gluon Plasma ?}

\begin{figure}[htbp]
\begin{center}
\resizebox*{12.6cm}{!}{\includegraphics{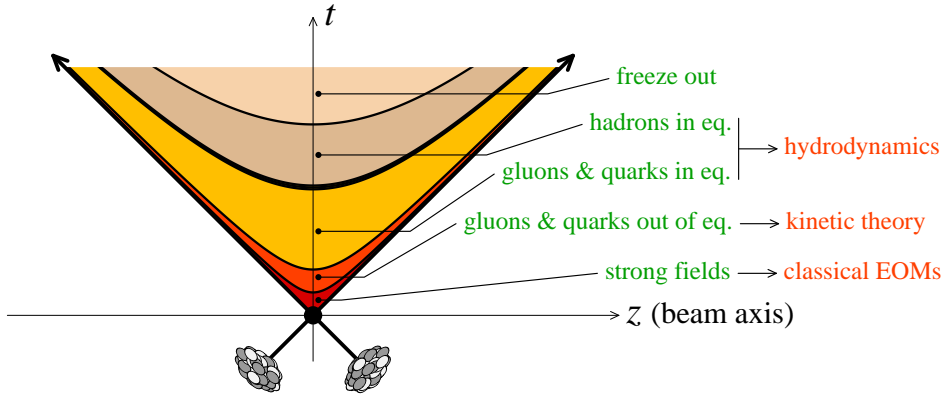}}
\end{center}
\caption{\label{fig:figII-00} Space-time development of a
  nucleus-nucleus collision. A goal of the Color Glass Condensate
  approach is to describe the first stage -- dominated by strong
  fields -- and to match it to the subsequent descriptions by kinetic
  theory or hydrodynamics.}
\end{figure}

In the previous lecture, we outlined a formalism to compute particle
production in field theories with strong time dependent sources. As
argued previously, the Color Glass Condensate is an example of such a
field theory. In the CGC framework, the high energy factorization
suggested by eq.~(\ref{eq:LI-0}) is assumed to compute final states.
In this lecture, we will address the question of how one computes in
practice the initial Glasma fields after a heavy ion collision, what
the properties of these fields are and outline theoretical approaches
to understanding how these fields may thermalize to form a Quark Gluon
Plasma. A cartoon depicting the various stages of the spacetime evolution 
of matter in a heavy ion collision is shown in fig.~\ref{fig:figII-00}.

In the CGC effective field theory, hard (large $x$) parton modes in
each of the nuclei are Lorentz contracted, static sources of color
charge for the soft (small $x$) wee parton, Weizs\"acker--Williams
modes in the nuclei.  Here $x$ is the longitudinal momentum fraction
of partons in the colliding nuclei. Wee modes with $x \ll A^{-1/3}$
and $ k_\perp \geq \Lambda_{_{\rm QCD}}$ are coherent across the
longitudinal extent of the nucleus and therefore couple to a large
density of color sources.  With increasing energy, the scale
separating soft and hard modes shifts towards smaller values of $x$;
how this happens can be quantified by a Wilsonian RG~\cite{IancuV1}.
In a heavy ion collision, the color current corresponding to the large
$x$ modes can be expressed as
\begin{equation}
J^{\mu,a}=\delta^{\mu +} \rho_{1}^a(\x_\perp) \delta(x^-)+
\delta^{\mu -} \rho_{2}^a(\x_\perp) \delta(x^+)\;,
\label{eq:LII-1}
\end{equation}
where the color charge densities $\rho_{1,2}^a$ of the two nuclei are
independent sources of color charge on the light cone. Let us recall
that $x^\pm=(t\pm z)/\sqrt{2}$. The $\delta$ functions represent the
fact that Lorentz contraction has squeezed the nuclei to
infinitesimally thin sheets. The absence of a longitudinal size scale
ensures that the gauge fields generated by these currents will be
boost-invariant -- they are independent of the space time rapidity
$\eta\equiv{\rm atanh}({z}/{t})$.

The gauge fields before the collision are obtained by solving the
Yang-Mills equations $$D_\mu F^{\mu \nu}=J^\nu\;,$$ where $D_\mu \equiv
\partial_\mu+i g [A_\mu,.]$ and $F_{\mu \nu}\equiv\partial_{\mu}
A_{\nu}-\partial_{\nu} A_{\mu}+i g [A_{\mu},A_{\nu}]$ are the gauge
covariant derivative and field strength tensor, respectively, in the
fundamental representation and $[A_\mu,.]$ denotes a commutator.

Before the nuclei collide ($x^0 < 0$), a solution of the equations of
motion is~\cite{McLerV1,McLerV2}
\begin{equation}
A^{\pm}=0 \quad,\quad
A^i= \theta_\epsilon(x^-)\theta_\epsilon(-x^+)\alpha_1^i(\x_\perp)
+
\theta_\epsilon(x^+)\theta_\epsilon(-x^-)
\alpha_2^i(\x_\perp) \; ,
\label{eq:LII-2}
\end{equation}
where, here and in the following, the transverse coordinates $x,y$ are
labeled by the Latin index $i=1,2$.  The subscript $\epsilon$ on the
$\theta$-functions denote that they are smeared by an amount
$\epsilon$ in the respective $x^\pm$ light cone directions. We require
that the functions $\alpha_m^i(\x_\perp)$ ($m=1,2$ denote the labels
of the colliding nuclei) are such that $F^{ij}=0$ -- they are pure
gauge solutions of the equations of motion. The gauge fields, just as
the Weizs\"acker--Williams fields in QED, are therefore plane
polarized sheets of radiation before the collision.  The functions
$\alpha_m^i$ satisfy
\begin{equation}
-D_i \alpha_{m}^i = \rho_{m}({\x_\perp}) \; .
\label{eq:LII-3}
\end{equation}
This equation has an analytical solution given
by~\cite{Kovch1,JalilKMW1}
\begin{equation}
\alpha^i_{m}=-\frac{i}{g}\; e^{i \Lambda_{m}}\; 
\partial^i\; e^{- i \Lambda_{m}}\quad, 
\qquad {\bs\nabla}^2_\perp \Lambda_{m}=-g\, \rho_{m}\; .
\label{eq:LII-4}
\end{equation}
To obtain this result one has to assume path ordering in $x^\pm$
respectively for nucleus 1 and 2; we assume that the limit
$\epsilon\rightarrow 0$ is taken at the end of the calculation.

We now introduce the proper time $\tau\equiv\sqrt{t^2-z^2}=\sqrt{2 x^+
  x^-}$ -- the initial conditions for the evolution of the gauge field
in the collision are formulated on the proper time surface $\tau =0$.
They are obtained~\cite{KovneMW1,KovneMW2} by generalizing the
previous ansatz for the gauge field to
\begin{eqnarray}
&&
A^i(x^-,x^+,x^\perp)=
\theta_\epsilon(x^-)\theta_\epsilon(-x^+) \; \alpha_1^i(\x_\perp)
+
\theta_\epsilon(-x^-) \theta_\epsilon(x^+) \; \alpha_2^i(\x_\perp)
\nonumber\\
&&\qquad\qquad\qquad\qquad\qquad
+\theta_\epsilon(x^-)\theta_\epsilon(x^+)\;
\alpha^i_3(x^-,x^+,\x_\perp)\nonumber\\
&&
A^{\pm}=\pm x^\pm \;\theta_\epsilon(x^-)\theta_\epsilon(x^+) \;
\beta(x^-,x^+,\x_\perp)\; ,
\label{eq:LII-5}
\end{eqnarray}
where we adopt the Fock--Schwinger gauge condition $A^\tau \equiv x^+
A^-+x^- A^+=0$. This gauge is an interpolation between the two light
cone gauges $A^\pm=0$ on the $x^\pm=0$ surfaces respectively.

The gauge fields $\alpha_3,\beta$ in the forward light cone can be
determined from the known gauge fields $\alpha_{1,2}$ of the
respective nuclei before the collision by invoking a physical
``matching condition'' which requires that the Yang-Mills equations
$D_\mu F^{\mu \nu}=J^\nu$ be regular at $\tau=0$. The
$\delta$-functions of the current in the Yang--Mills equations
therefore have to be compensated by identical terms in spatial
derivatives of the field strengths.  Interestingly, it leads to the
unique solution~\cite{KovneMW2}
\begin{eqnarray}
&&
\alpha_3^i(x^+,x^-,\x_\perp)=\alpha^i_1(\x_\perp)+\alpha^i_2(\x_\perp)
\nonumber\\
&& 
\beta(x^+,x^-,\x_\perp)=-\frac{ig}{2}\;
[\alpha_{i,1}(\x_\perp),\alpha_2^i(\x_\perp)]\; .
\label{eq:LII-6}
\end{eqnarray}
Further, the only condition on the derivatives of the fields that
would lead to regular solutions are $\partial_\tau
\beta |_{\tau=0}\;,\;\partial_\tau \alpha_3^i |_{\tau=0} =0$.

For the purpose of solving the Yang-Mills equations for a heavy-ion
collision on a lattice, we shall work with the $\tau,\eta$
co-ordinates and re-express the initial conditions for the fields and
their derivatives in terms of the fields and their conjugate momenta
in these co-ordinates. Our gauge condition is $A^\tau=0$, and the
initial conditions in eq.~(\ref{eq:LII-6}) for the functions
$\alpha_3^i$ and $\beta$ at $\tau=0$ can be expressed in terms of the
fields
\begin{eqnarray}
&&
 A_i(\tau,\x_\perp)\equiv A_i(\tau,\eta,\x_\perp) \; ,
\nonumber\\
&&
\Phi (\tau,\x_\perp) \equiv 
x^+ A^-(\tau,\eta,\x_\perp)
-
x^- A^+(\tau,\eta,\x_\perp)
\; ,
\label{eq:LII-7}
\end{eqnarray}
where we have made manifest the fact that these fields are
boost-invariant -- i.e. independent of $\eta$.  This is a direct
consequence of the assumption in eq.~(\ref{eq:LII-1}) that the
currents are $\delta$-function sources on the light cone.  The light
cone Hamiltonian in $A^\tau=0$ gauge, in this case of boost invariant
fields, can be written as~\cite{KrasnV1}
\begin{equation}
{\mathcal H}={\rm Tr}\ \left[\frac{E_i^2}{\tau}+\frac{(D_i \Phi)^2}{\tau}+\tau
E_\eta^2 + \tau F_{x y}^2\right] \, . 
\label{eq:LII-8}
\end{equation}
Here the conjugate momenta to the fields are the chromo-electric fields
\begin{equation}
E_i\equiv\tau\,
\partial_\tau A_i
\quad,\qquad 
E_\eta\equiv
\frac{1}{\tau}
\partial_\tau \Phi\; .
\label{eq:LII-9}
\end{equation}
Note that the contribution of the hard valence current does not appear
explicitly in the $A^\tau=0$ Hamiltonian expressed in ($\tau$, $\eta$)
co-ordinates. The dependence on the color source densities is entirely
contained in the dependence of the initial conditions on the source
densities.  Boost invariance simplifies the problem tremendously
because the QCD Hamiltonian in this case is ``dimensionally reduced"
to a $2+1$-d (QCD + adjoint scalar field) Hamiltonian.

In terms of these Glasma fields and their conjugate momenta, the
initial conditions in eq.~(\ref{eq:LII-6}) at $\tau=0$ can be
rewritten as
\begin{eqnarray}
&&
A^i(0,\x_\perp)=\alpha_{1}^i(\x_\perp)+\alpha^i_{2}(\x_\perp)\;, 
\nonumber\\
&&
\Phi(0,\x_\perp)=0\; ,
\nonumber \\
&&
E_i(0,\x_\perp)= 0\;, 
\nonumber\\
&&
E_\eta(0,\x_\perp)=i\, g\, [\alpha_{1}^i(\x_\perp),\alpha_{2}^i(\x_\perp)]\;.
\label{eq:LII-10}
\end{eqnarray}
The magnetic fields being defined as $B_k \equiv
\epsilon_{k\mu\nu}F^{\mu\nu}$, these initial conditions suggest that
$B_\eta \neq 0$ and $B_i =0$. Note that the latter condition follows
from the constraint on the derivatives of the gauge field that ensure
regular solutions at $\tau=0$. Thus one obtains the interesting
results that the initial Glasma fields correspond to large initial
longitudinal electric and magnetic fields ($E_\eta, B_\eta \neq 0$)
and zero transverse electric and magnetic fields ($E_i,B_i=0$). This
is in sharp contrast to the electric and magnetic fields of the nuclei
before the collision (the Weizs\"acker--Williams fields) which are
purely transverse!  Their importance was emphasized recently by Lappi
and McLerran~\cite{LappiMcLerran} who also coined the term ``Glasma"
to describe the properties of these fields prior to equilibration.

An immediate consequence of these initial conditions, as noted by
Khar\-zeev, Krasnitz and Venugopalan~\cite{KKV}, is that non-zero
Chern-Simons charge can be generated in these collisions. The dynamics
of the Chern-Simons number in nuclear collisions however differs from
the standard discussion in two ways. Firstly, the time translational
invariance of the fields is broken by the singularity corresponding to
the collision.  Secondly, due to the boost invariance of the
solutions, there can be no non-trivial boost invariant gauge
transformations. This can be seen as follows. In Ref.~\cite{KKV}, it
was shown that the Chern-Simons charge per unit rapidity could be
expressed as
\begin{equation}
\nu = \frac{1}{16\pi^2}\int d^2 \x_\perp\, \Phi^a B_\eta^a \; .
\label{eq:LII-11}
\end{equation}
Because this density is manifestly invariant under ra\-pi\-di\-ty
de\-pen\-dent trans\-for\-ma\-tions, such transformations (which
correspond to sphaleron transitions) cannot change the Chern-Simons
charge. Thus sphaleron transitions are disallowed for boost-invariant
field configurations.  Eqs.~(\ref{eq:LII-10}) tell us that
$\nu(\tau=0)=0$; therefore the Chern-Simons charge generated in a
given window in rapidity at a time $\tau$ is simply, by definition, $
\nu(\tau)(\eta_{\rm max}-\eta_{\rm min})$. Since $\eta$'s of either
sign are equally likely, the ensemble average $\langle
\nu(\tau)\rangle$ is zero.  However, $\langle \nu(\tau)^2\rangle$ is
non zero.  Its value was computed in Ref.~\cite{KKV}. The topological
charge squared per unit rapidity generated for RHIC and LHC collisions
is about 1-2 units.  In contrast, estimates of the same quantity in a
thermal plasma are one to two orders of magnitude larger. If boost
invariance is violated (as we shall soon discuss), sphaleron
transitions can go, and can potentially be large. This possibility, in
a different formulation, was discussed previously by Shuryak and
collaborators~\cite{Shuryak}.

We shall now discuss the particle distributions that correspond to the
gauge fields and their conjugate momenta in the forward light cone.
From the Hamilton equations
\begin{equation}
\frac{\partial {\mathcal H}}{\partial E_\mu}=\partial_\tau A_\mu
\quad,\quad
\frac{\partial {\mathcal H}}{\partial A_\mu}=-\partial_\tau E_\mu \; ,
\label{eq:LII-12}
\end{equation}
the Yang-Mills equations are 
\begin{eqnarray}
&&
\partial_\tau A_i=\frac{E_i}{\tau}\; ,
\nonumber\\
&&
\partial_\tau A_\eta =
\tau E_\eta\; ,
 \nonumber\\
&&
\partial_\tau E_i =\tau D_j F_{ji}+\tau^{-1} D_\eta F_{\eta i}\; ,
\nonumber\\
&&
\partial_\tau E_\eta =\tau^{-1}D_j F_{j\eta}\; .
\label{eq:LII-13}
\end{eqnarray}
They also satisfy the Gauss law constraint 
\begin{equation}
D_i E_i + D_\eta E_\eta =0 \; .
\label{eq:LII-14}
\end{equation}
These equations are non-linear and have to be solved numerically.  A
lattice discretization is convenient because it preserves gauge
invariance explicitly. One can write down the analogue of the well
known Kogut--Susskind Hamiltonian in this case and solve
eq.~(\ref{eq:LII-13}) numerically on a discretized
spatial\footnote{The proper time $\tau$ is treated as a continuous
  variable, that can have increments as small as required to reach the
  desired accuracy in the solution of the equations of motion.}
lattice with the initial conditions in eq.~(\ref{eq:LII-10}). We shall
not describe the numerical procedure here but instead refer the reader
to Refs.~\cite{KrasnV1,KrasnNV2}.

Solving Hamilton's equations, the average gluon multiplicity can be
computed using precisely the formula we discussed previously in
eq.~(\ref{eq:LI-26}). The result in eq.~(\ref{eq:LI-26}) is the
average multiplicity for {\it a} configuration of color charge
densities in each of the nuclei. It is an average in the sense of
being the first moment of the multiplicity distribution. This
multiplicity has to be further averaged over the distribution of
sources $W_{_{Y_{\rm beam}-Y}}[\rho_1]$ and $W_{_{Y_{\rm beam}+Y}}
[\rho_2]$, as specified in eq.~(\ref{eq:LI-0}).  These weight
functionals have to be specified at an initial scale $Y_0$ in rapidity
, and are then evolved to higher rapidities by the JIMWLK
renormalization group
equation~\cite{JalilKMW1,JalilKLW1,JalilKLW2,JalilKLW3,JalilKLW4,IancuLM1,IancuLM2,FerreILM1}.
For the purposes of computing the average multiplicity in {\it
  central} Au-Au collisions at RHIC, i.e. for rapidities where
evolution effects {\it a la JIMWLK} are not yet important, the weight
functionals $W_{_Y}[\rho]$ are Gaussian distributions specified in the
MV model (discussed briefly in the introduction to these lectures):
\begin{equation}
  W_{_{Y_0}}[\rho_{1,2}] = 
\exp\left( - \int d^2 \x_\perp \frac{\rho_{1,2}^a\rho_{1,2}^a}{2\, \Lambda_s^2} \right)\; ,
\label{eq:LII-15} 
\end{equation}
Here $\Lambda_s^2 = g^4 \mu^2$, where $g^2 \mu^2$ is the color charge
squared of the sources per unit area. The nuclei, for simplicity, are
assumed to be identical nuclei. $\Lambda_s^2$ is the only dimensionful
scale (besides the nuclear radius $R$) in the problem. It is simply
related, in leading order, to the nuclear saturation scale $Q_s$ by
the expression $\Lambda_s^2 = 2\pi Q_s^2/N_c\ln({Q_s}/{2 L})$, where
$L$ is an infrared scale of order $\Lambda_{\rm QCD}$. The nuclear
saturation scale, performing a simple extrapolation of the HERA data
on the gluon distribution of the proton to Au nuclei, is of the order
$Q_s\sim 1.2-1.4$ GeV at RHIC energies in several
estimates~\cite{KowalskiTeaney,KLN}.  For $L=0.2$ GeV, this
corresponds to a value $\Lambda_s = 1.66-1.8$ GeV. Clearly, there are
logarithmic uncertainties in this estimate at least of order 10\%.
For the rest of this lecture, we will assume the Gaussian form in
eq.~(\ref{eq:LII-15}) for the averaging over sources; modifications to
account for the (very likely) significant effects of small $x$ quantum
evolution will have to be considered at LHC energies.

\begin{figure}[htbp]
\begin{center}
\includegraphics[width=2.7in]{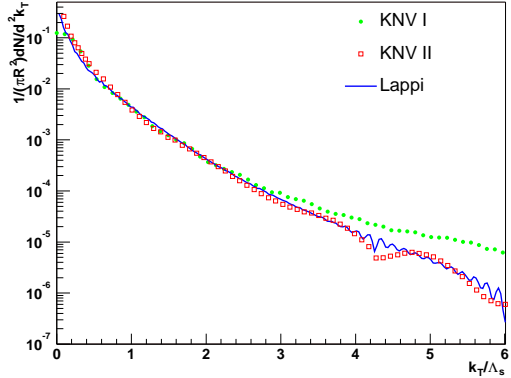}
\end{center}
\caption{
\label{fig:dndkt}Comparison of gluon transverse momentum distributions per
  unit area as a function of $k_{_T}/\Lambda_s$.  KNV I (circles):
  the number defined with $k_{_T}$ taken to mean the lattice wave
  number along one of the principal directions. KNV II (squares) and
  Lappi (solid line): the number defined by averaging over the entire
  Brillouin zone and with $k_{_T}$ taken to mean the frequency
  $\omega({k}_\perp)$. }
\end{figure}

In order to compute gluon number distributions, we impose the
transverse Coulomb gauge ${\bs\nabla}_\perp\cdot {\bs A}_\perp=0$ to
fix the gauge freedom completely.  The result for the number
distributions, averaged over the sources in eq.~(\ref{eq:LII-15}), is
computed at a time $\tau \sim 3/\Lambda_s$ to be
\begin{equation}
 \frac{1}{\pi R^2}\;\frac{dN}{d\eta d^2\k_\perp}
     = \frac{1}{g^2}\;{\overline f}_n(k_\perp/\Lambda_s)\; ,
\label{eq:LII-16}
\end{equation}
where ${\overline f}_n(k_\perp/\Lambda_s)$ is a function of the form
\begin{eqnarray}
{\overline f}_n = \left\{
  \begin{array}{ll}
     a_1\left[\exp\left(\sqrt{k_\perp^2+ m^2}/ T_{\rm eff}\right) -1\right]^{-1}
                       & (k_\perp/\Lambda_s \leq 1.5) \\ \\                

    a_2\,\Lambda_s^4\;\ln(4\pi k_\perp/\Lambda_s)\;k_\perp^{-4}
                       & (k_\perp/\Lambda_s > 1.5) \\
\end{array} \right.\; ,
\label{eq:LII-17}
\end{eqnarray}
with $a_1=0.137$, $m=0.0358\,\Lambda_s$, $T_{\rm
  eff}=0.465\,\Lambda_s$, and $a_2=0.0087$. These results are plotted
in fig.~\ref{fig:dndkt} -- they are compared to those computed
independently by Lappi~\cite{Lappi1}. The different lines in the
figure correspond to different lattice discretizations; the
differences at large $k_\perp$ therefore indicate the onset of lattice
artifacts, which can be eliminated by going closer to the continuum
limit (larger lattices).

\begin{figure}[htbp]
\begin{center}
\includegraphics[angle=-90,width=2.7in]{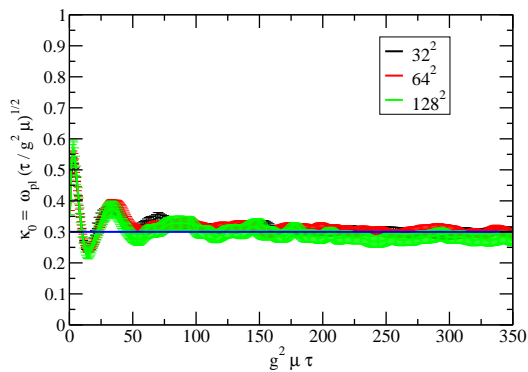}
\end{center}
\caption{
  \label{fig:plasmon_tau}Time evolution of the ``plasmon frequency''
  $\omega_{\rm pl}$, for fixed $g^2 \mu L=22.6$ and lattice spacings
  $g^2 \mu a_\perp=0.707,0.354,0.177$ ($N_\perp=32,64,128$),
  respectively.}
\end{figure}

From eq.~(\ref{eq:LII-17}), the number distribution at large $k_\perp$
has the power law dependence one expects in perturbative QCD at
leading order.  For small $k_\perp$, the result is best fit by a
massive 2-d Bose-Einstein distribution even though one is solving
classical equations of motion!  There is an interesting discussion in
the statistical mechanics literature that suggests that such a
distribution may be generic for classical ``glassy" systems far from
equilibrium~\cite{CaratiG}.  Another interesting observation is that
the non-perturbative real time dynamics of the gauge fields generates
a mass scale $m$ which makes the number distributions infrared safe
for finite times. Such a ``plasmon mass'' can be extracted from the
single particle dispersion relation; it behaves dynamically as a
function of time precisely as a screening mass
does~\cite{KrasnV2,PaulRaju1}. This can be seen in
fig.~\ref{fig:plasmon_tau}.  As we shall discuss shortly, this plasmon
mass can be related to the growth rate of instabilities in the Glasma.

The total transverse energy and number can be obtained independently,
from the Hamiltonian density and from a gauge invariant relaxation
(cooling) technique respectively. These agree with those obtained by
integrating eq.~(\ref{eq:LII-16}) over $k_\perp$ and can be expressed
as
\begin{equation}
\frac{1}{\pi R^2}\;\frac{dE_\perp}{d\eta} 
= \frac{f_{_E}(\Lambda_s R)}{g^2}\; \Lambda_s^3 
\quad,\qquad
\frac{1}{\pi R^2}\;\frac{dN}{d\eta} 
= \frac{f_{_N}(\Lambda_s R)}{g^2}\; \Lambda_s^2 \; ,
\label{eq:LII-18}
\end{equation}
where $f_{_E}=0.27-0.25$ and $f_{_N}=0.315-0.3$ for the wide range
$\Lambda_s R=50-167$ respectively. For larger values of $\Lambda_s
R$, the functions $f_{_E}$ and $f_{_N}$ have a weak logarithmic
dependence on $\Lambda_s R$.  If we assume parton-hadron duality and
directly compare the number of gluons from eq.~(\ref{eq:LII-18}) to
the number of hadrons measured at $\eta=0$ in $\sqrt{s} =
200$~GeV/nucleon Au-Au collisions at RHIC, one obtains a good
agreement for $\Lambda_s \approx 2$~GeV. This value is a little larger
than the values we extracted from extrapolations of the HERA data; one
should however keep in mind that additional contributions to the
multiplicity of hadrons will accrue from quark and gluon production at
next-to-leading order~\cite{GelisKL1}. If we include these contributions, as we hope to 
eventually, $\Lambda_s$ will be lower than this value.

The ``formation time'' $\tau_f$, defined as the time when the energy
density $\varepsilon$ behaves as $1/\tau$, is defined as $\tau_f =
1/\gamma\Lambda_s$, where $\gamma=0.3$ in the range of interest. The
initial energy density for times $\tau > \tau_f$ ($\tau_f \approx
0.3$~fm for $\Lambda_s=2$ GeV)is then
\begin{equation}
\varepsilon = \frac{1}{\tau_f}\;
\frac{dE_\perp}{{\pi R^2 d\eta}} = 
\frac{0.26}{g^2}\;\frac{\Lambda_s^3}{\tau} \; .
\label{eq:LII-19}
\end{equation}
This energy density, again for $\Lambda_s=2$~GeV (and $g=2$), is
$\varepsilon\approx 40$~GeV/fm$^3$ at $\tau=\tau_f$. Because the
energy density is ultraviolet sensitive, this number is probably an
overestimate because the spectrum at large $k_\perp > \Lambda_s$ in
practice falls much faster than the lowest order estimate in
eq.~(\ref{eq:LII-17}). In a recent paper~\cite{Lappi2}, Lappi has
shown that the energy density computed in this framework, at early
times has the form $\varepsilon \sim \ln^2(1/\tau)$; it is finite for
any $\tau > 0$ but is not well defined strictly at $\tau=0$.

In the discussion up to this point, we have assumed that the color
charge squared per unit area of the source, $g^2\mu^2$, is constant.
However, for finite nuclei, this is not true and one can define an
impact parameter dependent $\Lambda_s$, i.e. $\Lambda_s(\x_\perp)$.
This generalization, in the classical Yang-Mills framework described
here was discussed previously in Ref.~\cite{KrasnNV2} and is given by
 \begin{equation}
\Lambda_s^2 (\x_\perp) = \Lambda_{s0}^2 \,T_{_A}(\x_\perp)\; ,
\label{eq:LII-20} 
\end{equation} 
where $T_{_A}(\x_\perp) = \int^{\infty}_{-\infty} dz\,
\rho_{_{WS}}(z,\x_\perp)$ is the nuclear thickness profile, $\x_\perp$
is the transverse coordinate vector (the reference frame here being
the center of the nucleus), $\rho_{_{WS}}(z,\x_perp)$ is the
Woods-Saxon nuclear density profile, and $\Lambda_{s0}^2$ is the color
charge squared per unit area in the center of the nucleus. One can use
this expression to compute the multiplicity as a function of impact
parameter in the collision. Then, by using a Glauber model to relate
the average impact parameter to the average number of
participants~\cite{KharzeevN}, one can obtain the dependence of the
multiplicity on the number of participants.

Previous computations of the centrality dependence of the multiplicity
and of rapidity distributions were performed in the KLN
approach~\cite{KLN,KharzeevN,KharzeevL}. There however, unlike
eq.~(\ref{eq:LII-20}), the saturation scale depends on the number of
participant nucleons:
\begin{equation} 
Q_{s}^2(\x_\perp) \sim 
{\rm N_{part}}(\x_\perp)\; , 
\label{eq:LII-21}
\end{equation} 
with 
\begin{equation} {\rm N_{part}}(\x_\perp) 
= 
T_{_A}(\x_\perp) \;
 \left[
1- \left(1-\sigma_{_{NN}} 
\frac{T_{_B}(\x_\perp-\b)}{B} \right)^B \right]\; .
\label{eq:LII-22}
\end{equation}
In this formula, $\sigma_{_{NN}}$ is the nucleon-nucleon
cross-section, and $\b$ the impact parameter between the two nuclei.
Note that as this form involves the thickness functions of both nuclei
$A$ and $B$, it is manifestly not universal -- in contrast to the definition in 
eq.~\ref{eq:LII-20}. For the
centrality dependence of the multiplicity distributions, the
saturation scales defined through eqs.~(\ref{eq:LII-20}) or
(\ref{eq:LII-21}) lead to very similar results. This is because the
multiplicity, at any particular $\x_\perp$, depends on the lesser of
the two saturation scales, say $Q_{s,_A}$. The dependence on the
``non-universal'' factor in eq.~(\ref{eq:LII-22}) is then weak
because, by definition, $T_{_B}$ is large.

\begin{figure}[htbp]
\begin{center}
\includegraphics[width=2.7in]{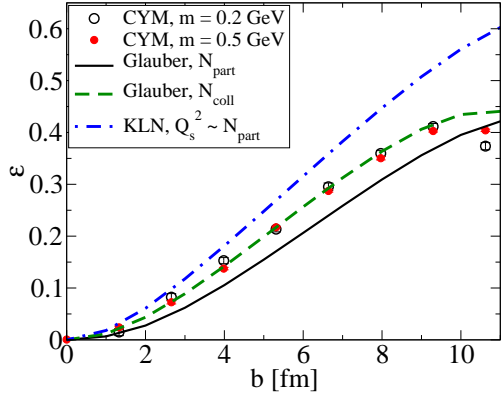}
\end{center}
\caption{The eccentricity as a function of impact parameter. The
  classical field CGC result with two different infrared cutoffs $m$
  is denoted by CYM~\cite{LappiV}. The traditional initial
  eccentricity used in hydrodynamics is a linear combination of mostly
  ``Glauber ${\rm N_{part}}$'' and a small amount of ``Glauber ${\rm
    N_{coll}}$''. The ``KLN'' curve is the eccentricity obtained from
  the CGC calculation in Refs.~\cite{Hirano:2005xf,Drescher:2006pi}. }
\label{fig:ecc}
\end{figure}

However, the two prescriptions can be distinguished by examining a
quantity of phenomenological importance, the eccentricity
$\epsilon$ defined as
\begin{equation}
\epsilon \equiv
\frac
{\int d^2 \x_\perp\, \varepsilon(\x_\perp)\, (y^2 - x^2) }
{\int d^2 \x_\perp\, \varepsilon(\x_\perp)\, (y^2 + x^2) }\; .
\label{eq:LII-23}
\end{equation}
This quantity is a measure of the asymmetry of the overlap region
between the two nuclei in collisions at non zero impact parameter. In
an ideal hydrodynamical description of heavy ion collisions, a larger
initial eccentricity may lead to larger elliptic
flow~\cite{Ollitrault:1992bk} than observed, thereby necessitating
significant viscous effects. Comparisons of model predictions, with
different initial eccentricities, to data may therefore help constrain
the viscosity of the Quark Gluon Plasma.  In fig.~\ref{fig:ecc}, we
show results for the eccentricity from the $k_\perp$ factorized KLN
approach with the saturation scale defined as in eq.~(\ref{eq:LII-21})
compared to the classical Yang-Mills (CYM) result computed with the
definition in eq.~(\ref{eq:LII-20}).
The KLN ${\rm N_{part}}$ definition of $Q_s$ leads to the largest
eccentricity.  The universal CYM definition gives smaller values of
$\epsilon$ albeit larger than the traditional parameterization (used in
hydrodynamical model computations) where the energy density is taken
to be proportional to the number of participating nucleons. This
result is also shown to be insensitive to two different choices of the
infrared scale $m$ which regulates the spatial extent of the Coulomb
tails of the gluon distribution. A qualitative explanation of the
differences in the eccentricity computed in the two approaches is
given in Ref.~\cite{LappiV} -- we refer the reader to the discussion
there. Also, a further elaboration of the discussion of Refs.~\cite{Hirano:2005xf,Drescher:2006pi} was 
very recently presented in Ref.~\cite{DrescherNara}--the revised curves our closer to the result in Ref.~\cite{LappiV}.

Our discussion thus far of the Glasma has assumed strictly boost
invariant initial conditions on the light cone, of the form specified
in eq.~(\ref{eq:LII-10}). However, this is clearly an idealization
because it requires strict $\delta$-function sources as in
eq.~(\ref{eq:LII-1}), {\sl and} that one completely disregards quantum
fluctuations. Because the collision energy is ultra-relativistic and
because quantum fluctuations are suppressed by one power of
$\alpha_s$, this was believed to be a good approximation. In
particular, it was not realized that violations of boost invariance
lead to a non-Abelian version of the Weibel instability~\cite{Weibel}
well known in electromagnetic plasmas. To understand the potential
ramifications of this instability for thermalization, let us first
consider where the boost invariant results lead us. From
eq.~(\ref{eq:LII-19}), it is clear that the energy density is far from
thermal -- in which event, it would decrease as $\tau^{-4/3}$. The
momentum distributions become increasingly anisotropic: $\langle
k_\perp \rangle \sim Q_s$ and $\langle k_z\rangle \sim 1/\tau$. Once
the particle-like modes of the classical field ($k_\perp > \Lambda_s$)
begin to scatter, the occupation number of the field modes begins to
decrease. How this occurs through scattering was outlined in an
elegant scenario dubbed ``bottom up'' by Baier et al~\cite{BMSS}. At
very early times, small elastic scattering of gluons with $k_\perp
\sim Q_s$ dominates and is responsible for lowering the gluon
occupation number. The Debye mass $m_{_D}^2 \sim Q_s^2/(Q_s\tau)$ (see
fig.~\ref{fig:plasmon_tau}) sets the scale for these scattering, and
the typical $k_z$ is enhanced by collisions. One obtains $k_z \sim
Q_s/(Q_s\tau)^{1/3}$. From this dependence, one can estimate that the
occupation number of gluons is $f <1$ for $\tau >
\alpha_s^{-3/2}/Q_s$. For proper times greater than these, the
classical field description becomes less reliable. In the bottom up
scenario, soft gluon radiation from $2\rightarrow 3$ scattering
processes becomes important at $\tau\sim \alpha_s^{-5/2}/Q_s$. The
system thermalizes shortly thereafter at $\tau_{\rm therm} =
\alpha_s^{-13/5}/Q_s$ with a temperature $T\sim \alpha_s^{2/5} Q_s$.
The thermalization time scale in this scenario is parametrically
faster than that obtained by solving the Boltzmann equation for
$2\rightarrow 2$ processes, which gives $\tau_{\rm therm} \sim
\exp(\alpha_s^{-1/2})/Q_s$~\cite{Mueller2000,BjorakerV}~\footnote{For
  recent numerical studies of thermalization due to scattering, see
  Ref.~\cite{GreinerXu}.}.

The Debye mass scale is key to the power counting in the bottom up
scenario. However, as pointed out recently~\cite{ALM} this power
counting is affected~\footnote{For a modified ``bottom up'' scenario,
  we refer the reader to Ref.~\cite{MuellerShoshiWong}.} by an
instability that arises from a change in sign of the Debye mass
squared for anisotropic momentum
distributions~\cite{RomatschkeStrickland}. The instability is the
non-Abelian analog of the Weibel instability~\cite{Weibel} in
electromagnetic plasmas and was discussed previously in the context of
QCD plasmas by Mrowczynski~\cite{Mrowczynski}. One can view the
instability, in the configuration space of the relevant fields, as the
development of specific modes for which the effective potential is
unbound from below~\cite{AL}. Detailed simulations in the Hard-Loop
effective theory in $1+1$-dimensions~\cite{ALM,AL,RRS1} and in
$3+1$-dimensions~\cite{AMY,RRS2} have confirmed the existence of this
non-Abelian Weibel instability.  Particle--field simulations of the
effects of the instability on thermalization have also been performed
recently~\cite{DumitruNara,DumitruNaraStrickland,Schenke-etal}.

All of these simulations consider the effect of instabilities in
systems at rest. However, as discussed previously, the Glasma expands
into the vacuum at nearly the speed of light. Are they seen in the
Glasma ? No such instabilities were seen in the boost invariant
$2+1$-d numerical simulations.  In the rest of this lecture, we will
discuss the consequences of relaxing boost invariance in (now) $3+1$-d
numerical simulations of the Glasma fields~\footnote{This discussion
  of $3+1$-d numerical simulations is based on work in
  Refs.~\cite{PaulRaju1,PaulRaju2}}; as may be anticipated,
non-Abelian Weibel instabilities also arise in the Glasma.

In heavy-ion collisions, the initial conditions on the light cone are
never exactly boost invariant. Besides the simple kinematic effect of
Lorentz contraction at high energies, one also has to take into
account quantum fluctuations at high energies.  For instance, as we
discussed in the last lecture, we will have small quantum fluctuations
at NLO, for each configuration of the color sources, which are not
boost invariant.  Parametrically, from the power counting discussed
there, quantum fluctuations may be of order unity, compared to the
leading classical fields which are of order $1/g$.

In the following, we will discuss two simple models of initial
conditions containing rapidity dependent fluctuations.  A more
complete theory should specify, from first principles, the initial
conditions in the boost non-invariant case. We will discuss later some
recent work in that direction. The only condition we impose is that
these initial conditions satisfy Gauss' law.  We construct these by
modifying the boost-invariant initial conditions in
eq.~(\ref{eq:LII-10}) to
\begin{eqnarray}
&&
E_i(0,\eta,\x_\perp)=\delta E_i(\eta,\x_\perp)\; ,
\nonumber\\
&&
E_\eta(0,\eta,\x_\perp)=i\, g\, [\alpha_{1}^i,\alpha_{2}^i]+
  \delta E_\eta(\eta,\x_\perp,)\; ,
\label{eq:LII-24}
\end{eqnarray}
while keeping $A_i,A_\eta$ unchanged. The rapidity dependent
perturbations $\delta E_i,\delta E_\eta$ are in principle arbitrary,
except for the requirement that they satisfy the Gauss law. For these
initial conditions, it takes the form
\begin{equation}
D_i \delta E_i+D_\eta E_\eta=0 \; .
\label{eq:LII-25}
\end{equation}
The boost invariance violating perturbations are constructed as
follows.
\begin{itemize}
\item  We first generate random configurations 
$\delta \overline{E}_i(\x_\perp)$ with 
$$\langle
\delta \overline{E}_i(\x_\perp) 
\delta \overline{E}_j(\y_\perp)
\rangle =
\delta_{ij} \delta(\x_\perp-\y_\perp)\; .$$
\item Next, for our first model of rapidity perturbations, we generate
a Gaussian random function $F(\eta)$ with amplitude $\Delta$
\begin{equation}
\langle F(\eta) F(\eta^\prime)\rangle=\Delta^2 \delta(\eta-\eta^\prime)\;.
\label{eq:LII-26}
\end{equation}
For the second model, we also generate a Gaussian random function, but
subsequently remove high-frequency components of $F(\eta)$
\begin{equation}
F(\eta)\rightarrow F(\eta)=\int \frac{d\nu}{2\pi}\; e^{-i \nu \eta} \;
e^{-|\nu| b}
\int d\eta\;
e^{i \eta \nu}\; F(\eta)\; ,
\label{eq:LII-27}
\end{equation}
where $b$ acts as a ``band filter'' suppressing the high frequency
modes. This model is introduced because the white noise Gaussian
fluctuations of the previous model leads to identical amplitudes for
all modes.  As a consequence, the high momentum modes dominate bulk
observables such as the pressure. The unstable modes we wish to focus
on are sensitive to infrared modes at early times but their effects
are obscured by the higher momentum modes from the white noise
spectrum. This is particularly acute for large violations of boost
invariance. Therefore, damping these high frequency modes allows us to
also study the effect of instabilities for larger values of $\Delta$,
or ``large seeds'' that violate boost-invariance.
\item For both models, once $F(\eta)$ is generated, we obtain for the
fluctuation fields
\begin{eqnarray}
&&
\delta E_i(\eta,\x_\perp)=
\partial_\eta F(\eta) 
\delta\overline{E}_i(\x_\perp)\; ,
\nonumber\\
&&
\delta E_\eta(\eta,\x_\perp)=-F(\eta) D_i \delta
\overline{E}_i(\x_\perp) \; .
\label{eq:LII-28}
\end{eqnarray}
\end{itemize}
These fluctuations, by construction, satisfy Gauss' law. To implement
rapidity fluctuations in the above manner, one requires $\tau>0$. This
is a consequence of the $\tau,\eta$ coordinates\footnote{This system
of coordinates becomes singular when $\tau\to 0$, as can be seen from
the fact that the Jacobian for the transformation from Cartesian
coordinates vanishes in this limit.} and does not have a physical
origin. We therefore implement these initial conditions for
$\tau=\tau_{\rm init}$ with $0<\tau_{\rm init}\ll Q_s^{-1}$. Our
results are only weakly dependent on the specific choice of $\tau_{\rm
init}$.

The primary gauge invariant observables in simulations of the
classical Yang-Mills equations are the components of the
energy-momentum tensor~\cite{KrasnV1,KrasnNV2}. We will discuss
specifically
\begin{eqnarray}
&&
T^{xx}+T^{yy}=2\, {\rm Tr}\left[F^2_{xy}+E_\eta^2\right]\; ,
\nonumber\\
&&
\tau^2 T^{\eta \eta}=\tau^{-2}\ {\rm Tr}\left[F_{\eta i}^2+E_i^2\right]
-{\rm Tr}\left[F_{xy}^2+E_\eta^2\right]\;.
\label{eq:LII-30}
\end{eqnarray}
Note that the Hamiltonian density is ${\mathcal H}=\tau T^{\tau
\tau}\equiv \tau (T^{xx} + T^{yy} + \tau^2 T^{\eta\eta})$.  These
components can be expressed as
\begin{equation}
\tau P_\perp = \frac{\tau}{2} \left(T^{xx}+T^{yy}\right)\quad, \qquad
\tau P_{_L}=\tau^3 T^{\eta \eta}\;,
\label{eq:LII-31}
\end{equation}
which correspond to $\tau$ times the mean transverse and longitudinal
pressure, respectively.

When studying the time evolution of rapidity-fluctuations, it is
useful to introduce Fourier transforms of observables with respect to
the rapidity. For example,
\begin{equation}
\widetilde{P_{_L}}(\tau,\nu,\k_\perp=0)\equiv
\int d \eta \;e^{i \eta\nu}\; 
\langle P_{_L}(\tau,\eta,\x_\perp)\rangle_\perp\,,
\label{eq:LII-32}
\end{equation}
where $\langle \rangle_\perp$ denotes averaging over the transverse
coordinates $(x,y)$. Apart from $\nu=0$, this quantity would be
strictly zero in the boost-invariant ($\Delta=0$) case, while for
non-vanishing $\Delta$ and $\nu$, $\tilde{P_{_L}}(\nu)$ has a maximum
amplitude for some specific momentum $\nu$. Using a very small but
finite value of $\Delta$, this maximum amplitude is very much smaller
than the corresponding amplitude of a typical transverse momentum
mode.

The physical parameters in this study are $g^2\mu$ (=$\Lambda_s$; see
the discussion after eq.~(\ref{eq:LII-15})), $L^2= \pi R^2$, where $R$
is the nuclear radius, $\Delta$, the initial size of the rapidity
dependent fluctuations and finally, the band filter $b$, which as
discussed previously, we employ only for large values of $\Delta$.
Physical results are expressed in terms of the dimensionless
combinations $g^2\mu \tau$ and $g^2\mu L$. For RHIC collisions of gold
nuclei, one has $g^2\mu L \approx 120$; for collisions of lead nuclei
at LHC energies, this will be twice larger. The physical properties of
the spectrum of fluctuations (specified in our simple model here by
$\Delta$ and $b$) will presumably be further specified in a complete
theory. For our present purposes, they will be treated as arbitrary
parameters, and results presented for a large range in their values.

Briefly, the lattice parameters in this study, in dimensionless units,
are (i) $N_\perp$ and $N_\eta$, the number of lattice sites in the
$\x_\perp$ and $\eta$ directions respectively; (ii) $g^2\mu\,a_\perp$
and $a_\eta$, the respective lattice spacings; (iii) $\tau_0/a_\perp$
and $\delta\tau$, the time at which the simulations are initiated and
the stepping size respectively. The continuum limit is obtained by
holding the physical combinations $g^2\mu\,a_\perp\,N_\perp = g^2 \mu
L$ and $a_\eta\,N_\eta = L_\eta$ fixed, while sending $\delta \tau$,
$g^2\mu\,a_\perp$ and $a_\eta$ to zero.  For this study, we pick
$L_\eta = 1.6$ units of rapidity.  The magnitude of violations of
boost invariance, as represented by $\Delta$, is physical and deserves
much study. The initial time is chosen to ensure that for $\Delta=0$,
we recover earlier $2+1$-d results; we set $\tau_0
=0.05\,a_\perp$. Our results are insensitive to variations that are a
factor of 2 larger or smaller than this choice.  For further details
on the numerical procedure we refer the reader to
Refs.~\cite{PaulRaju1,PaulRaju2}.
\begin{figure}[htbp]
\begin{center}
\includegraphics[angle=-90,width=2.7in]{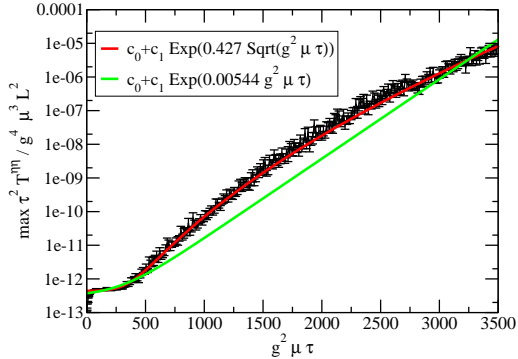}
\end{center}
\caption{
\label{fig:maxFM}The maximum Fourier mode amplitudes of the mean longitudinal
pressure $P_L = \tau^2 T^{\eta\eta}$ for $g^2\mu L=67.9$,
$N_\perp=N_\eta=64$, $N_\eta a_\eta=1.6$.  Also shown are best fits
with $\exp(a\tau)$ and $\exp(a\sqrt{\tau})$ behavior. The former is
clearly ruled out by the data.}
\end{figure}

\begin{figure}[htbp]
\begin{center}
\includegraphics[width=2.7in]{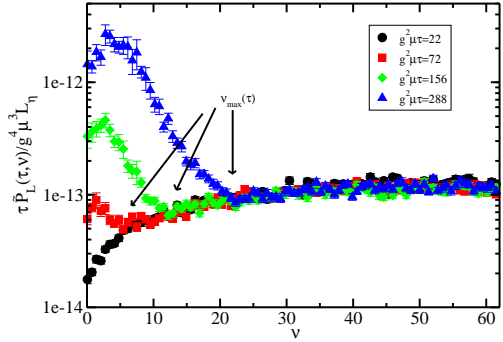}
\end{center}
\caption{
\label{fig:earlytimes}
$\tau \widetilde{P}_{_{L}}(\tau, \nu)$ as a function of momentum
$\nu$, averaged over 160 initial conditions on a $16^2\times2048$
lattice with $g^2 \mu L=22.6$ and $L_\eta=102.4$, $\Delta\simeq
10^{-11}$. Four different simulation times show how the softest modes
start growing with an distribution reminiscent of results from
Hard-Loop calculations \cite{RomatschkeStrickland}. Also indicated are
the respective values of $\nu_{\rm max}$ for three values of $g^2 \mu
\tau$ (see text for details).}
\end{figure}

In fig.~\ref{fig:maxFM}, we plot the maximal value\footnote{What we
mean by this is further clarified by fig.~\ref{fig:earlytimes} and the
related discussion there.} of $\tau^2
\widetilde{T}^{\eta\eta}=\widetilde{P}_{_L}$ at each time step, as a
function of $g^2\mu \tau$. The data are for a $64^3$ lattice and
correspond to $g^2\mu L = 67.9$ and $L_\eta = 1.6$. The maximal value
remains nearly constant until $g^2\mu \tau \approx 250$, beyond which
it grows rapidly. A best fit to the functional form $c_0 + c_1
\exp({c_2 \tau^{c_3}})$ gives $c_2 =0.427\pm 0.01$ for $c_3 = 0.5$;
the coefficients $c_0$, $c_1$ are small numbers proportional to the
initial seed. It is clear from fig.~\ref{fig:maxFM} that the form
$\exp(\Gamma \sqrt{g^2\mu \tau})$ is preferred to a fit with an
exponential growth in $\tau$.  This $\exp(\Gamma \sqrt{g^2\mu \tau})$
growth of the unstable soft modes is closely related to the mass
generated by the highly non-linear dynamics of soft modes in the
Glasma.  As we discussed previously, and showed in
fig.~\ref{fig:plasmon_tau}, a plasmon mass $\omega_{\rm pl} \equiv
\omega(\k_\perp =0)$, is generated.  After an initial transient
behavior, it is of the form
\begin{equation}
\omega_{\rm pl} = \kappa_0 \,g^2\mu \,\sqrt{\frac{1}{g^2\mu\tau}}\;.
\label{eq:LII-33}
\end{equation}
with $\kappa_0 = 0.3\pm 0.01$ (this parameterization is robust as one
approaches the continuum limit). The dependence on $g^2\mu L$ is weak.

In the finite temperature Hard Thermal Loop (HTL) formalism for
anisotropic plasmas, the maximal unstable modes of the stress-energy
tensor grow as $\exp(2\gamma\tau)$), where the growth rate $\gamma$
satisfies the relation $\gamma=m_\infty/\sqrt{2}$ for maximally
anisotropic particle distributions~\cite{ALM}. Here
\begin{equation}
m_\infty^2 =   g^2 N_c\,\int \frac{d^3 \p}{(2\pi)^3} 
\frac{f(\p)}{p}\;,
\label{eq:LII-34}
\end{equation}
where $f(\p)$ is the anisotropic single particle distribution of the
hard modes.  It was shown in Ref.~\cite{RomatschkeStrickland} that
$m_\infty^2 = 3\omega_{\rm pl}^2/2$ for both isotropic and
a\-ni\-so\-tro\-pic plasmas. One therefore obtains
\begin{equation}
2\,\gamma\,\tau = \sqrt{3}\frac{\kappa_0\, g^2\mu }{\sqrt{g^2\mu\tau}}\tau 
=
\sqrt{3}\kappa_0\sqrt{g^2\mu\,\tau}\;.
\label{eq:LII-35}
\end{equation}
 For $g^2\mu L = 67.9$, $\kappa_0 \approx 0.26$ gives the coefficient
$\sqrt{3}\kappa_0\approx 0.447$, which is quite close to the value
obtained by a fit to the numerical data $\Gamma_{\rm fit}
=0.427$. However, this agreement is misleading because a proper
treatment would give in the exponent $\int^\tau d\tau^\prime
\gamma(\tau^\prime) = \Gamma_{\rm th} \sqrt{g^2\mu \tau}$, with
$\Gamma_{\rm th} = 2 \sqrt{3}\,\kappa_0$.  The observed growth rate is approximately 
half of that predicted by directly applying the HTL formalism to the
Glasma. Despite obvious similarities, it is not clear that the
equivalence can be expected to hold at this level of
accuracy. Nevertheless, the similarities in the two frameworks is
noteworthy as we will discuss now.

In fig.~\ref{fig:earlytimes} we show the ensemble-averaged $\tau
\widetilde{P}_{_L}(\tau,\nu)$ for four different simulation times.
The earliest time ($g^2 \mu \tau\simeq 22$) shows the configuration
before the instability sets in. At the next time, one sees a bump
above the background, corresponding to the distribution of unstable
modes. The unstable mode with the biggest growth rate (the cusp of the
``bumps'' in fig.\ref{fig:earlytimes}) was precisely what was used to
determine the maximal growth rate $\Gamma$ by fitting the time
dependence of this mode to the form $c_0+c_1 \exp{(\Gamma \sqrt{g^2
\mu \tau})}$. The two later time snapshots shown in
fig.~\ref{fig:earlytimes} (for $g^2 \mu \tau\simeq 156$ and $g^2 \mu
\tau\simeq 288$) indicate that the growth rate of the unstable modes
closely resembles the analytic prediction from Hard-Loop
calculations~\cite{ALM,RomatschkeStrickland}.

\begin{figure}[htbp]
\begin{minipage}[t]{.45\linewidth}
\begin{center}
\includegraphics[width=\linewidth]{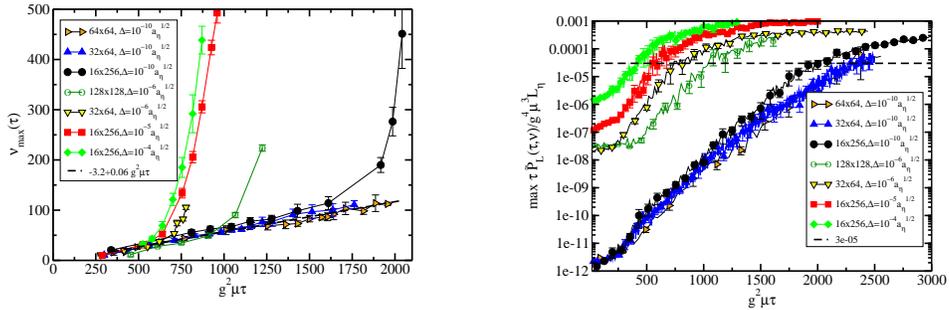}
\end{center}
\end{minipage}
\hfill
\begin{minipage}[t]{.45\linewidth}
\begin{center}
\includegraphics[width=\linewidth]{maxamp.eps}
\end{center}
\end{minipage}
\caption{Left: Time evolution of $\nu_{\rm max}$, on a lattices with
$g^2 \mu L=22.7$, $L_\eta=1.6$ and various violations of
boost-invariance $\Delta$. The dashed line represents the linear
scaling behavior. Right: Time evolution of the maximum amplitude $\tau
\widetilde{P}_{_L}(\tau,\nu)$. When this amplitude reaches a certain size
(denoted by the dashed horizontal line), $\nu_{\rm max}$ starts to
grow fast.}
\label{fig:numax}
\end{figure}

In fig.~\ref{fig:earlytimes}, $\nu_{\rm max}$ is the largest mode
number that is sensitive to the instability. Its behavior is shown in
fig.~\ref{fig:numax}.
From this figure, one observes an underlying trend indicating a linear
increase of $\nu_{\rm max}$ with approximately $\nu_{\rm max}\simeq
0.06\, g^2 \mu \tau$. For sufficiently small violations of
boost-invariance, this seems to be fairly independent of the
transverse or longitudinal lattice spacing we have tested. For much
larger violations of boost-invariance -- or sufficiently late times --
one observes that $\nu_{\rm max}$ deviates strongly from this ``linear
law''. In fig.~\ref{fig:numax} we show that this deviation seems to
occur when the maximum amplitude of $\tau \tilde P_L(\tau,\nu)$
reaches a critical size, independent of other simulation
parameters. This critical value is denoted by a dashed horizontal line
and has the magnitude $3\cdot 10^{-5}$ in the dimensionless units
plotted there.  A possible explanation for this behavior is that once
the transverse magnetic field modes in the Glasma (with small
$k_\perp$) reach a critical size, the corresponding Lorenz force in
the longitudinal direction is sufficient to bend ``particle'' (hard
gauge mode) trajectories out of the transverse plane into the
longitudinal direction. This is essentially what happens in
electromagnetic plasmas. Note however that in electromagnetic plasmas
the particle modes are the charged fermions.In contrast, the particle
modes here are the hard ultraviolet transverse modes of the field
itself. We will comment shortly on how this phenomenon may impact
thermalization.

\begin{figure}[htbp]
\begin{center}
\includegraphics[width=2.7in]{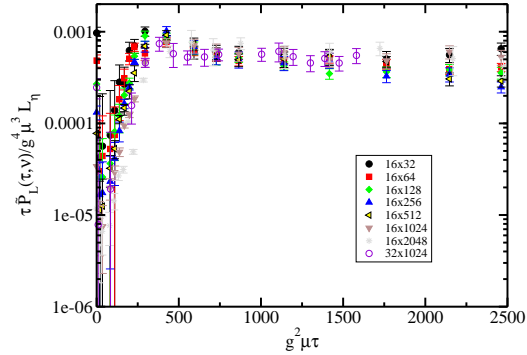}
\end{center}
\setlength{\unitlength}{1cm}
\caption{Time evolution of the (ensemble-averaged) maximum amplitude
of $\tau \widetilde{P}_{_L}(\tau,\nu)$, for $g^2\mu L=22.6$, $L_\eta=1.6$,
$N_\perp=16,32$, $\Delta=0.1\, a_\eta^{1/2}$ and $N_\eta$ ranging from
$32$ to $2048$. Larger lattices correspond to smaller $\Delta$.  This
explains why the early-time behavior is not universal for the
simulations shown here.}
\label{fig:LSsat}
\end{figure}

The saturation seen in the right of fig.~\ref{fig:numax} is shown
clearly in fig.~\ref{fig:LSsat} where we plot\footnote{Note here that
  for the case of ``large seed'' instabilities, we consider the model
  with the ``band filter'' $b=0.5$, in order to suppress the
  ultraviolet modes in the initial fluctuation. Note further that for
  larger seeds the instability systematically saturates at earlier
  times, as is clear from the right of fig.~\ref{fig:numax}.}
the temporal evolution of the maximum amplitude of the ensemble
averaged $\tau \tilde{P_L}(\tau,\nu)$, for lattices with different
$a_\eta$. Early times in this figure ($g^2 \mu \tau<200$) correspond
to the stage when the Weibel instability is operative.  Interestingly,
the simulations show saturation of the growth at approximately the
same amplitude. These preliminary results are similar to the
phenomenon of ``non-Abelian saturation'', found in the context of
simulations of plasma instabilities in the Hard Loop
framework~\cite{AMY,RRS2}.

\begin{figure}[htbp]
\begin{minipage}[t]{.45\linewidth}
\begin{center}
\includegraphics[width=\linewidth]{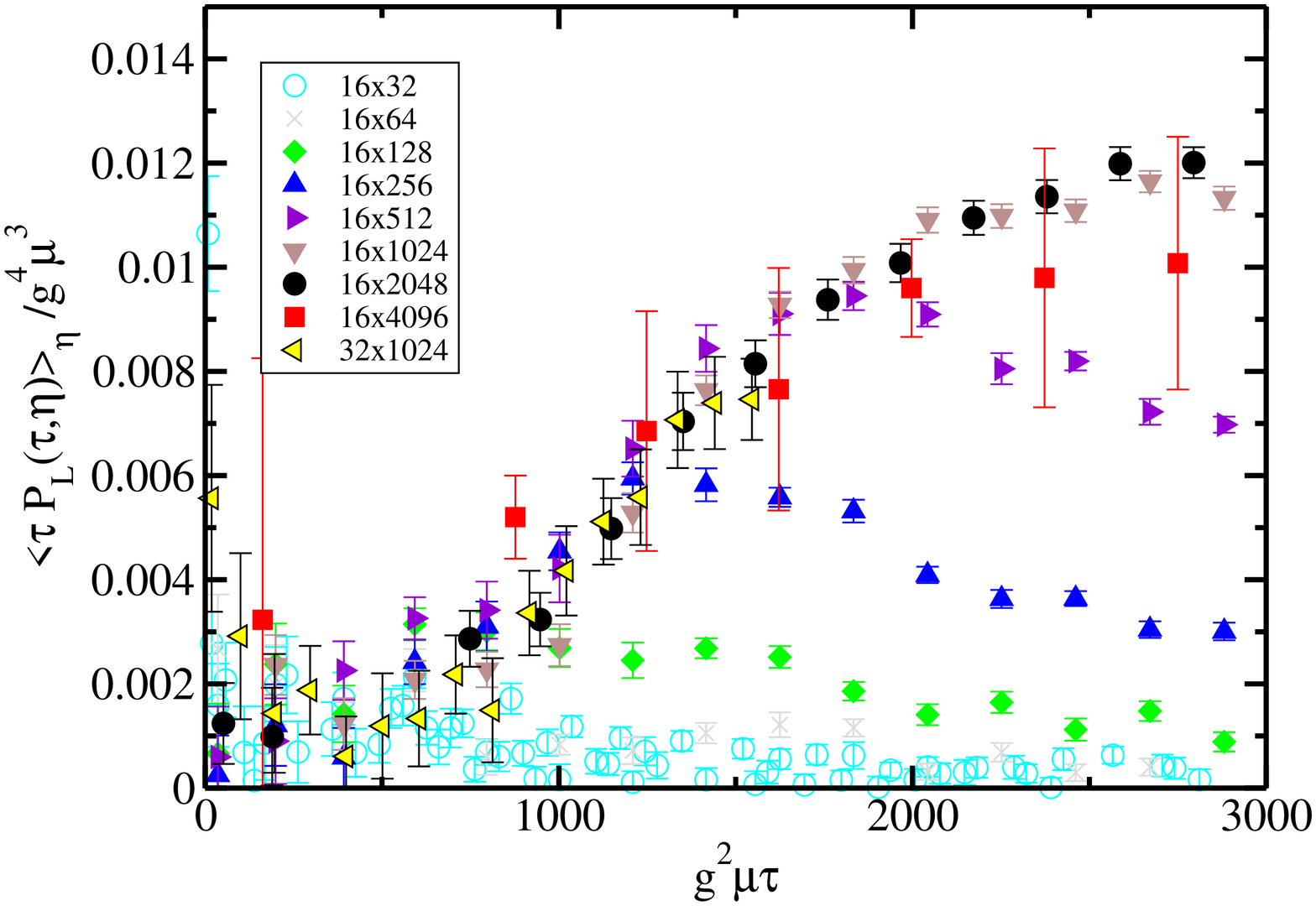}
\end{center}
\end{minipage}
\hfill
\begin{minipage}[t]{.45\linewidth}
\begin{center}
\includegraphics[width=\linewidth]{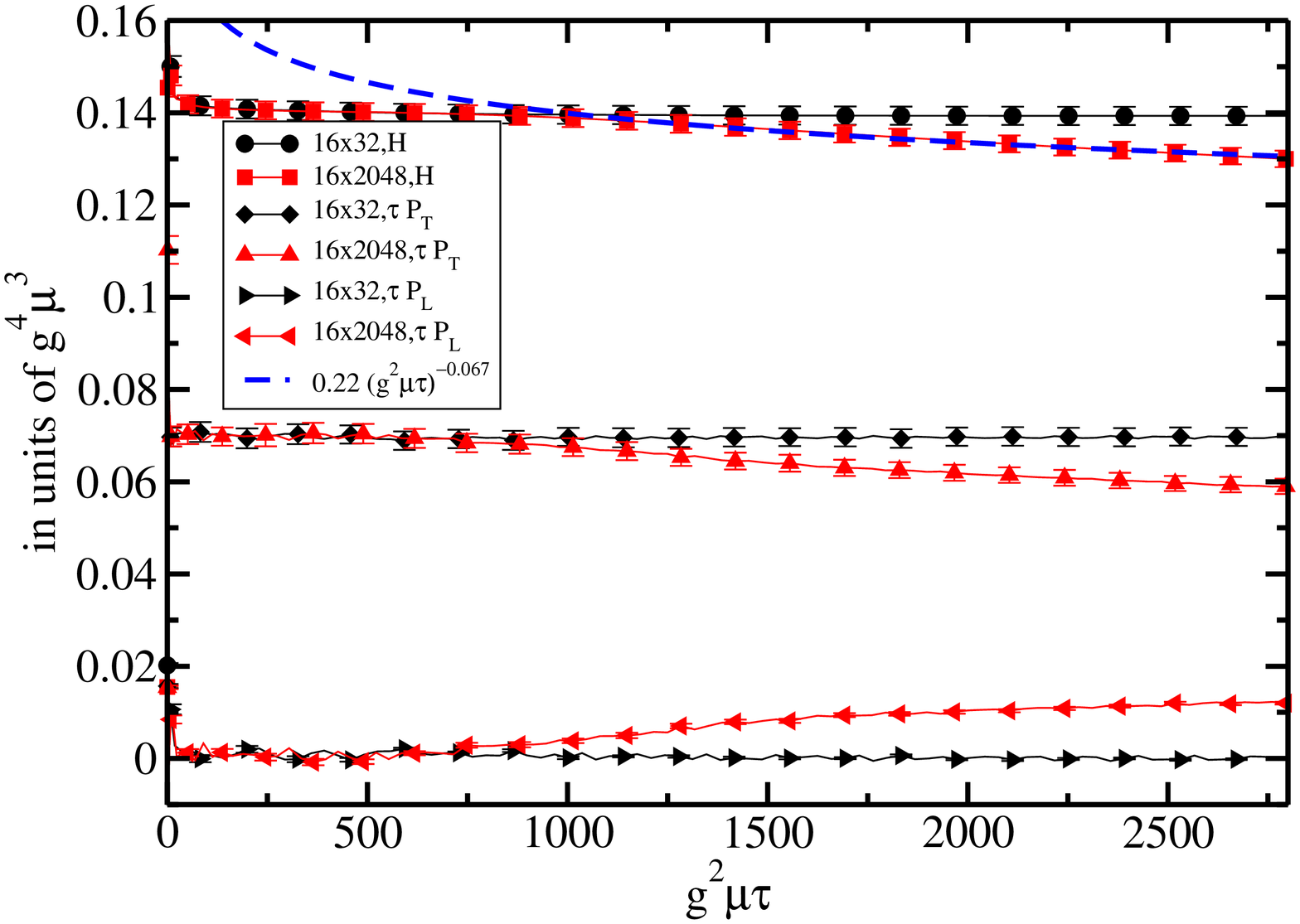}
\end{center}
\end{minipage}
\caption{Left: Time evolution of the ensemble and volume averaged
longitudinal pressure $P_{_L}$, for lattices with $g^2\mu L=22.6$,
$L_\eta=1.6$, $N_\perp=16,32$, $\Delta=0.1 a_\eta^{1/2}$ and $N_\eta$
ranging from $32$ to $2048$. Note~: reduced statistical ensemble of 2
runs for $N_\eta=4096$. Right: Hamiltonian density $\langle
H(\tau,\eta)\rangle_\eta$, $\tau \langle
P_{_T}(\tau,\eta)\rangle_\eta$ and $\tau \langle
P_{_L}(\tau,\eta)\rangle_\eta$, for $16^2\times32$ and $16^2\times2048$
lattices.  The energy density is fit to $\varepsilon \sim
\tau^{-1.067}$ at late times. All curves are calculated on lattices
with $g^2 \mu L=22.6$, $L_\eta=1.6$ and $\Delta=0.1 a_\eta^{1/2}$.}
\label{fig:Tetaeta}
\end{figure}

In the small seed case, the longitudinal fluctuations carry a tiny
fraction of the total system energy. In the large seed case, in
contrast, for the simulations shown here, the initial energy contained
in the longitudinal modes is $\sim 1$\% of the total system energy. In
reality, we expect this fraction to be significantly larger. However,
this would require us to study the dynamics on even larger
longitudinal lattices than those included in this study to ensure that
the contributions to the pressure from ultraviolet modes are not
contaminated by lattice artifacts. In the left fig.~\ref{fig:Tetaeta},
we plot $P_{_L}$ as a function of $\tau$ for different lattice
spacings $a_\eta$. For large $a_\eta$ (low lattice UV cutoff), the
longitudinal pressure is consistent with zero; it is clearly finite
when the lattice UV cutoff is raised. However, the rise saturates as
there is no notable difference between the simulations for the three
smallest values of the lattice spacing.  At face value, this result
suggests that the rise in the longitudinal pressure is physical and
not a discretization artifact. Clearly, further studies on larger
transverse lattices are needed to strengthen this claim.

In the right of fig.~\ref{fig:Tetaeta}, we investigate the time
evolution of the transverse pressure and the energy density for (i) a
simulation with a low UV cutoff ($16^2\times32$ lattice) and (ii) a
simulation with a high UV cutoff ($16^2\times2048$ lattice). We observe that the rise in the mean longitudinal
pressure accompanies a drop both in the mean transverse pressure and
energy density. This result is consistent with the previously
advocated physical mechanism of the Lorenz force bending transverse UV
modes (thereby decreasing the transverse pressure) into longitudinal UV
modes (simultaneously raising the longitudinal pressure), thereby
pushing the system closer to an isotropic state. The energy density
depends on the proper time as $\varepsilon \sim \tau^{-1.067}$, which,
while not the free streaming result of $\varepsilon \sim \tau^{-1}$,
is also distinct from the $\varepsilon \sim \tau^{-4/3}$ required for
a locally isotropic system undergoing one dimensional expansion.
Furthermore, the time scales (noting that for RHIC energies $g^2\mu
\equiv \Lambda_s \approx 2$ GeV) are much larger than the time scales
of interest for early thermalization of the Glasma into a QGP. Similar
results were obtained in an analytical model of the late time behavior
of expanding anisotropic fields in the Hard Loop
formalism~\cite{PaulToni}.

Nevertheless, these simulations are proof in principle that
non-trivial dynamics can take place in the Glasma driving the system
towards equilibrium.  A mode analysis along the lines of that
performed recently by B\"{o}deker and Rummukainen~\cite{Dietrich} is
required to understand this dynamics in greater detail. In particular,
it would be useful to understand whether the rapid shift of unstable
modes to the ultraviolet (as seen in Fig.~\ref{fig:numax}) is due to a
turbulent Kolmogorov cascade as discussed in
Refs.~\cite{AM,Mueller-Kolm}. The most important task however is to
understand from first principles the spectrum of initial fluctuations
that break boost invariance. A first step in this direction was taken
in Ref.~\cite{FukushimaGM}. This issue is closely related to the NLO
computation of small fluctuations outlined in Lecture I.  At NLO, some
of these quantum fluctuations are accompanied by large logs in $x$.
Thus for $\alpha_s Y\sim 1$, these effects are large. To completely
understand which contributions from the small fluctuations can be
absorbed in the evolution of the initial wavefunctions~\footnote{Note
  that the weight functionals describing the distribution of color
  sources, in this case, will have a dependence on rapidity described
  by the JIMWLK RG equations. This implies that even if the underlying
  classical fields are nearly boost invariant, the spectrum of gluons
  obtained by averaging over all color configurations, are not boost
  invariant. One expects significant deviations from boost invariance
  when rapidity varies by an amount $\delta Y\sim 1/\alpha_s$.}, and to
isolate the remainder that contributes to the spectrum of initial
fluctuations, requires that we demonstrate factorization for inclusive
multiplicities. This work is in progress~\cite{GelisLV}. Finally,
another interesting problem is whether one can match the temporal
evolution of Glasma fields into kinetic equations at late times.  Such
a matching was considered previously in Ref.~\cite{MuellerSon}. The
early time strong field dynamics may however modify the power counting
assumed in these studies -- this possibility is also under active
investigation~\cite{GelisJV}. In conclusion, understanding the early
classical field dynamics of the Glasma and its subsequent
thermalization is crucial to understand how and when the system
thermalizes to form a QGP. The phenomenological implications of these
studies are significant because they influence the initial conditions
for hydrodynamic models. One such example that we discussed is the
initial eccentricity of the QCD matter; its magnitude may be relevant
for our understanding of just how ``perfect'', the perfect fluid
created at RHIC is.

\section*{Lecture III: Limiting Fragmentation in the CGC framework}

In the first two lectures, we discussed the problem of multi-particle
production for hadronic collisions where the large $x$ modes are
strong sources $\rho_{1,2}\sim 1/g$. This is a good model of the
dynamics in proton-proton collisions at extremely high energies or in
heavy ion collisions already at somewhat lower energies~\footnote{This
follows from the large density of color charges in the transverse
plane of the Lorentz contracted heavy nuclei.}. In
eq.~(\ref{eq:LI-0}), one has $\rho_{1,2}/k_\perp^2\sim 1$, where
$k_\perp$ is the typical momentum of the partons in the nuclei. As we
then discussed in lectures I and II, there is no small parameter in
the expansion in powers of these sources and one has to solve
classical equations of motion numerically to compute the average
inclusive multiplicities for gluon and quark production. However, for
asymmetric collisions, the most extreme example of which are
collisions of protons with heavy nuclei, one has a situation where
$\rho_p/k_\perp^2 \ll 1$ and $\rho_A/k_\perp^2 \sim 1$. The other
situation where a similar power counting is applicable is when one
probes forward (or backward) rapidities in proton-proton or
nucleus-nucleus collisions. In these cases, one is probing large $x$
parton distributions in one hadron (small color charge density --
$\rho_1/k_\perp^2\ll 1$) and small $x$ parton distributions in the
other (large color charge density -- $\rho_2/k_\perp^2 \sim 1$). In
these situations, analytical computations are feasible in the CGC
framework. In this lecture, we will discuss the phenomenon of limiting
fragmentation in this framework.

The hypothesis of limiting fragmentation~\cite{BCYY} in high energy
hadron-hadron collisions states that the pseudo-rapidity distribution
$\frac{dN}{d\eta'}$ (where $\eta'\equiv\eta-Y_{\rm beam}$ is the
pseudo-rapidity shifted by the beam rapidity $Y_{\rm
beam}\equiv\ln(\sqrt{s}/m_{\rm p})$) becomes independent of the
center-of-mass energy $\sqrt{s}$ in the region around $\eta'\sim 0$,
i.e.
\begin{equation}
\frac{dN_{\rm ch}}{d\eta'}(\eta',\sqrt{s},b)
=
\frac{dN_{\rm ch}}{d\eta'}(\eta',b)\; ,
\label{eq:LIII-1}
\end{equation}
where $b$ is the impact parameter.

Limiting fragmentation appears to have a wide regime of validity. It
was confirmed experimentally in $p\bar{p},pA,\pi A$ and $AA$
collisions at high energies \cite{SPS,FERMILAB,BRAHMS,PHOBOS}. More
recently, the BRAHMS and PHOBOS experiments at the Relativistic Heavy
Ion Collider (RHIC) at Brookhaven National Laboratory (BNL) performed
detailed studies of the pseudo-rapidity distribution of the produced
charged particles $dN_{\rm ch}/d\eta$ for a wide range ($-5.4 < \eta <
5.4$) of pseudo-rapidities, and for several center-of-mass energies
($\sqrt{s_{_{NN}}}=19.6, 62.4, 130 \mbox{\ and\ } 200 \, {\rm GeV}$)
in nucleus-nucleus (Au-Au and Cu-Cu) and deuteron-nucleus (d-Au)
collisions. Results for pseudo-rapidity distributions have also been
obtained over a limited kinematic range in pseudo-rapidity by the STAR
experiment at RHIC~\cite{STAR}.  These measurements have opened a new
and precise window on the limiting fragmentation phenomenon.

It is worth noting that this scaling is in strong disagreement with
boost invariant scenarios which predict a fixed fragmentation region
and a broad central plateau extending with energy.  It would therefore
be desirable to understand the nature of hadronic interactions that
lead to limiting fragmentation, and the deviations away from it. In a
recent article, Bialas and Je\.zabek~\cite{BialasJezabek}, argued that
some qualitative features of limiting fragmentation can be explained
in a two-step model involving multiple gluon exchange between partons
of the colliding hadrons and the subsequent radiation of hadronic
clusters by the interacting hadrons.  Here we will discuss how the
limiting fragmentation phenomenon arises naturally within the CGC
approach~\footnote{Our discussion is based on our
paper with A. Stasto~\cite{GelisSV}.} -- we shall address its relation to the
Bialas-Je\.zabek model briefly later.

Inclusive gluon production in proton-nucleus collisions was first
computed in Refs.~\cite{KovMueller,DumitruMcLerran}, and shown to be
$k_\perp$ factorizable in Ref.~\cite{KKT}. In Ref.~\cite{BlaizGV1},
the gluon field produced in pA collisions was computed explicitly in
Lorentz gauge $\partial_\mu A^\mu=0$. More recently, the gluon field
was also determined explicitly in the $A^+=0$ light-cone
gauge~\cite{GelisMT}. The inclusive multiplicity distribution of
produced gluons\footnote{For more complicated final states, like
  quark-antiquark pairs, $k_\perp$ factorization is explicitly
  violated \cite{BGV2,FujiiGV,KovTuchin,Schaefer}.} can be expressed in the $k_\perp$-factorized form as
~\cite{GriboLR1,KKT,BlaizGV1},
\begin{equation}
\frac{dN_{\rm g}}{dy d^2\p_\perp}
 = 
\frac{\alpha_s S_{_{AB}}}{2\pi^4 C_{_F} (\pi R_{_A}^2)(\pi R_{_B}^2)}\,
\frac{1}{p_\perp^2}
\int \frac{d^2 \k_\perp}{(2\pi)^2}\;
  \phi_{_A}(x_1,k_\perp) \,\phi_{_B}(x_2,\big|\p_\perp-\k_\perp\big|) \; .
\label{eq:LIII-2}
\end{equation}
The formula, as written here, is only valid at zero impact parameter
and assumes that the nuclei have a uniform density in the transverse
plane; the functions $\phi_{_{A,B}}$ are defined for the entire
nucleus.  $S_{_{AB}}$ denotes the transverse area of the overlap
region between the two nuclei, while $\pi R_{_{A,B}}^2$ are the total
transverse area of the nuclei, and $C_{_F} \equiv (N_c^2 -1)/ 2 N_c$
is the Casimir in the fundamental representation.

The longitudinal momentum fractions $x_1$ and $x_2$ are defined as 
\begin{equation}
x_1 \equiv \frac{p_\perp}{m_{\rm p}}\, e^{y-Y_{\rm beam}}\; ,\quad x_2
\equiv \frac{p_\perp}{m_{\rm p}}\,e^{-y-Y_{\rm beam}} \; ,
\label{eq:LIII-3}
\end{equation}
where $Y_{\rm beam}= \ln(\sqrt{s}/m_{\rm p})$ is the beam rapidity,
$m_{\rm p}$ is the proton mass, and $\p_\perp$ is the transverse
momentum of the produced gluon. The kinematics here is the
$2\rightarrow 1$ eikonal kinematics, which provides the leading
contribution to gluon production in the CGC picture.

The functions $\phi_{_A}$ and $\phi_{_B}$ are obtained from the
dipole-nucleus cross-sections for nuclei $A$ and $B$ respectively,
\begin{equation}
  \phi_{_{A,B}}(x,k_\perp)\equiv
  \frac{\pi R_{_{A,B}}^2 k_\perp^2}{4\alpha_s N_c} 
\int d^2\x_\perp\; e^{i \k_\perp \cdot \x_\perp } \; 
\left<{\rm Tr}\left(U^{\dagger}(0) U(\x_\perp)\right)\right>_{_Y} \; ,
\label{eq:LIII-4}
\end{equation}
where $Y\equiv\ln(1/x)$ and where the matrices $U$ are adjoint Wilson
lines evaluated in the classical color field created by a given
partonic configuration of the nuclei $A$ or $B$ in the infinite
momentum frame. For a nucleus moving in the $-z$ direction, they are
defined to be
\begin{eqnarray}
U(\x_\perp)\equiv T_+ \exp\left[-ig^2\int\limits_{-\infty}^{+\infty}
dz^+ \frac{1}{{\bs\nabla}_{\perp}^2}\,\rho(z^+,\x_\perp)\cdot T\right] \; .
\label{eq:LIII-5}
\end{eqnarray}
Here the $T^a$ are the generators of the adjoint representation of
$SU(N_c)$ and $T_+$ denotes the ``time ordering'' along the
$z^+$ axis. $\rho_a(z^+,\x_\perp)$ is a certain configuration of the
density of color charges in the nucleus under consideration, and the
expectation value $\big<\cdots\big>$ corresponds to the average over
these color sources $\rho_a$. 

As discussed previously, in the McLerran-Venugopalan (MV)
model~\cite{McLerV1,McLerV2,McLerV3}, where no quantum evolution
effects are included, the $\rho$'s have a Gaussian distribution, with
a 2-point correlator given by 
$$\langle \rho_a(0)\rho_b(\x_\perp)\rangle= \mu_{_A}^2
\delta_{ab}\delta^{(2)}(\x_\perp-\y_\perp)\; ,$$ where $\mu_{_A}^2
\equiv {A}/{2\pi R_{_A}^2}$ is the color charge squared per unit
area.  This determines $\phi_{_{A,B}}$ completely~\cite{KKT,BlaizGV1},
since the 2-point correlator is all we need to know for a Gaussian
distribution.  We will shortly discuss the small $x$ quantum evolution
of the correlator on the r.h.s. of eq.~(\ref{eq:LIII-4}).

These distributions $\phi_{_{A,B}}$, albeit very similar to the
canonical unintegrated gluon distributions in the hadrons, should not
be confused with the latter~\cite{KKT,BlaizGV1}. However, at large
$k_\perp$ ($k_\perp \gg Q_s$), they coincide with the usual
unintegrated gluon distribution. Note that the unintegrated gluon
distribution here is defined such that the proton gluon distribution,
to leading order satisfies
$$xG_p(x,Q^2) = \frac{1}{4\pi^3}\,\int_0^{Q^2} dl_\perp^2
\phi_p(x,l_\perp)\; .$$ 

From eq.~(\ref{eq:LIII-2}), it is easy to see how limiting
fragmentation emerges in the limit where $x_2\ll x_1$. In this
situation, the typical transverse momentum $k_\perp$ in the projectile
at large $x_1$ is much smaller than the typical transverse momentum
$\big|\p_\perp-\k_\perp\big|$ in the other projectile, because these
are controlled by saturation scales evaluated respectively at $x_1$
and at $x_2$ respectively. Therefore, at sufficiently high energies,
it is legitimate to approximate
$\phi_{_B}(x_2,\big|\p_\perp-\k_\perp\big|)$ by
$\phi_{_B}(x_2,p_\perp)$. Integrating the gluon distribution over
$\p_\perp$, we obtain
\begin{eqnarray}
\frac{dN_{\rm g}}{dy}
&=&
\frac{\alpha_s S_{_{AB}}}{2\pi^4 C_{_F} (\pi R_{_A}^2)(\pi R_{_B}^2)}\,
\int \frac{d^2\p_\perp}{p_\perp^2}\phi_{_B}(x_2,p_\perp)
\int \frac{d^2 \k_\perp}{(2\pi)^2}\;
\phi_{_A}(x_1,k_\perp)
\nonumber\\
&=&
\frac{S_{_{AB}}}{\pi^3 R_{_A}^2}
\int\frac{d^2 \k_\perp}{(2\pi)^2}\;
\phi_{_A}(x_1,k_\perp) \simeq \frac{S_{_{AB}}}{\pi R_{_A}^2} x_1g(x_1,\mu^2\simeq Q_s^2(x_2))\; .
\label{eq:LIII-6}
\end{eqnarray}
This expression is nearly independent of $x_2$ and therefore depends
only weakly on on $y-Y_{\rm beam}$.  To obtain the second line in the
above expression, we have used eq.~(\ref{eq:LIII-4}) and the fact that
the Wilson line $U$ is a unitary matrix. Therefore, details of the
evolution are unimportant for limiting fragmentation, only the
requirement that the evolution equation preserves unitarity. The
residual dependence on $x_2$ comes from the the upper limit $\sim
Q_s^B(x_2)$ of the integral in the second line.  This ensures the
applicability of the approximation that led to the expression in the
second line above. The integral over $\k_\perp$ gives the integrated
gluon distribution in the projectile, evaluated at a resolution scale
of the order of the saturation scale of the target. Therefore, the
residual dependence on $y+Y_{\rm beam}$ arises only via the scale
dependence of the gluon distribution of the projectile. This residual
dependence on $y+Y_{\rm beam}$ is very weak at large $x_1$ because it
is the regime where Bjorken scaling is observed.

The formula in eq.~(\ref{eq:LIII-6}) was used previously in
Ref.~\cite{Jamal}. The nuclear gluon distribution here is determined
by global fits to deeply inelastic scattering and Drell-Yan data. We
note that the glue at large $x$ is very poorly constrained at
present~\cite{NPD}. The approach of Bialas and Jezabek
\cite{BialasJezabek} also amounts to using a similar formula, although
convoluted with a fragmentation function (see eqs.~(1), (4) and (5) of
\cite{BialasJezabek} -- in addition, both the parton distribution and
the fragmentation function are assumed to be scale independent in this
approach). We will discuss the effect of fragmentation functions later
in our discussion.

Though limiting fragmentation can be simply understood as a
consequence of unitarity in the high energy limit, what may be more
compelling are observed deviations from limiting fragmentation and how
these vary with energy. We will now see whether deviations from
limiting fragmentation can be understood from the renormalization
group (RG) evolution of the unintegrated gluon distributions in
eq.~\ref{eq:LIII-2}. In particular, we study the RG evolution of these
distributions given by the Balitsky-Kovchegov (BK)
equation~\cite{Balit1,Kovch3}.

The BK equation is a non-linear evolution equation\footnote{It is
equivalent to the corresponding JIMWLK
equation~\cite{JalilKMW1,JalilKLW1,JalilKLW2,JalilKLW3,JalilKLW4,IancuLM1,IancuLM2,FerreILM1}
of the Color Glass Condensate, in a mean field (large $N_c$ and large
$A$) approximation where higher order dipole correlators are
neglected.} in rapidity $Y=\ln(1/x)$ for the forward scattering
amplitude $T(r_\perp,Y)$ of a {\sl quark-antiquark dipole} of size
$r_\perp$ scattering off a target in the limit of very high
center-of-mass energy $\sqrt{s}$ where $T$ is defined as:
\begin{equation}
 T(r_\perp,Y)= 1- \frac{1}{N_c}{\rm Tr} \left< {\widetilde
U}^\dagger(0) {\widetilde U}(\r_\perp)\right>_{_Y} \; .
 \label{eq:LIII-7}
 \end{equation}
 Here ${\widetilde U}$ is the corresponding Wilson line for the
 scattering of a quark-anti-quark dipole in the {\it fundamental}
 representation.  The correlators $U$ in eq.~(\ref{eq:LIII-4}), which
 are in the {\it adjoint} representation are Wilson lines for the
 scattering of a gluon dipole on the same target instead. The BK
 equation captures essential features of high energy scattering. When
 $ r_\perp \ll 1/Q_s$, one has color transparency; for $r_\perp \gg
 1/Q_s$, the amplitude $T\rightarrow 1$, and one obtains gluon
 saturation~\footnote{$Q_s$ is defined in terms of the requirement
 that $T={1}/{2}$ for $r_\perp = {2}/{Q_s}$.}. It is therefore
 an excellent model to study both limiting fragmentation as well as
 deviations from it.

It is convenient to express the BK equation in momentum space in terms
of the Bessel-Fourier transform of the amplitude
\begin{equation}
\widetilde{T}(k_\perp,Y)
\equiv
\int\limits_0^{+\infty} \frac{dr_\perp}{r_\perp}  \, 
J_0(k_\perp r_\perp) \,  T(r_\perp,Y)\; .
\label{eq:LIII-8}
\end{equation}
One obtains 
\begin{equation}
\frac{\partial {\widetilde{T}}(k_\perp,Y)}{\partial Y}
=
\overline{\alpha}_s (K\otimes \widetilde{T})(k_\perp,Y)
-
\overline{\alpha}_s \widetilde{T}^2(k_\perp,Y)\; ,
\label{eq:LIII-9}
\end{equation}
where we denote $\overline{\alpha}_s\equiv \alpha_s N_c/\pi$.  The
operator $K$ is the well known BFKL kernel in momentum
space~\cite{BalitL1,KuraeLF1}.

In the large $N_c$ and large $A$ limit, the correlators of Wilson
lines in the fundamental and adjoint representations are simply
related: $$\frac{1}{N_c}\;{\rm Tr}\langle
U(0)U^\dagger(\r_\perp)\rangle_{_Y}=N_c
\left[1-T(r_\perp,Y)\right]^2\; .$$ One can therefore express the
unintegrated gluon distribution in eq.~(\ref{eq:LIII-4}) in terms of
$T$ as
\begin{equation}
\phi_{_{A,B}}(x,k_\perp) 
= 
\frac{\pi^2 R_{_A}^2 N_c k_\perp^2}{2\,\alpha_s}
\int\limits_0^{+\infty} r_\perp dr_\perp 
\;
J_0(k_\perp r_\perp)
\,
\left[1-T_{_{A,B}}(r_\perp,\ln(1/x))\right]^2
 \; .
 \label{eq:LIII-10}
\end{equation}

\begin{figure}[htbp]
\begin{center}
\includegraphics[width=2.7in]{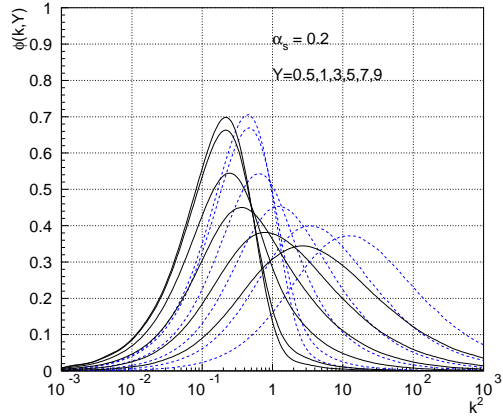}
\end{center}
\caption{The unintegrated gluon distribution from (i) correlators in
the adjoint representation (eq.\ref{eq:LIII-10}) (dashed lines) and
(ii) correlators in the fundamental representation -- see text (solid
lines).}
\label{fig:1a}
\end{figure}

In Ref.~\cite{GelisSV} we solved the BK equation numerically, in both
fixed and running coupling cases, in order to investigate limiting
fragmentation in hadronic collisions~\footnote{The BK equation was
solved numerically previously for various
studies~~\cite{ArmestoBraun,LevLub,GBMS,Albacete2,WeiRum}}. The
results are shown in fig.~\ref{fig:1a}.
The solid line is the result obtained for the unintegrated
distribution corresponding to correlators in the fundamental
representation, i.e. proportional to the Fourier transform of $1-T$
instead of that of $N_c (1-T)^2$ in eq.~(\ref{eq:LIII-10}).

Our results for limiting fragmentation are obtained through the
following procedure:
\begin{itemize}
\item One first solves the BK equation in eq.~(\ref{eq:LIII-9}) to
obtain (via eq.~(\ref{eq:LIII-8})) eq.~(\ref{eq:LIII-10}) for the
unintegrated distributions $\phi(x,k_\perp)$. The solution is performed
for $x \leq x_0=10^{-2}$ with the initial condition
$\phi(x_0,k_\perp)$, given by the McLerran-Venugopalan
model~\cite{McLerV1,McLerV2,McLerV3} with a fixed initial value of the
saturation scale $Q_s^A(x_0)$. For a gold nucleus, extrapolations from
HERA and estimates from fits to RHIC data suggest that
$(Q_s^A(x_0))^2\approx 2 \, {\rm GeV}^2$. The saturation scale in the
proton is taken to be $Q_s^2(x_0)= (Q_s^A(x_0))^2 \,(197/A)^{1/3}$.
For comparison, we also considered initial conditions from the
Golec-Biernat and Wusthoff (GBW) model~\cite{GBW}. The values of
$Q_s^A$ were varied in this study to obtain best fits to the data.
\item We used the ansatz $$\phi(x,k_\perp)=
\bigg(\frac{1-x}{1-x_0}\bigg)^{\beta} \,
\bigg(\frac{x_0}{x}\bigg)^{\lambda_0} \, \phi(x_0,k_\perp)\; ,$$ in
order to extrapolate our results to larger values of $x > x_0$, where
the parameter $\beta=4$ is fixed by QCD counting rules.
\item The resulting expressions are substituted in
eq.~(\ref{eq:LIII-2}) to determine rapidity distribution of the
produced gluons. The pseudo-rapidity distributions are determined by
multiplying eq.~(\ref{eq:LIII-2}) with the Jacobian for the
transformation from $y$ to $\eta$. This transformation requires one to
specify an infrared mass, which is also the mass chosen to regulate
the (logarithmic) infrared sensitivity of the rapidity distributions.
\end{itemize}
For further details on how the results are obtained, we refer the
reader to Ref.~\cite{GelisSV}.

\begin{figure}[hbt]
\begin{center}
\includegraphics[width=4in]{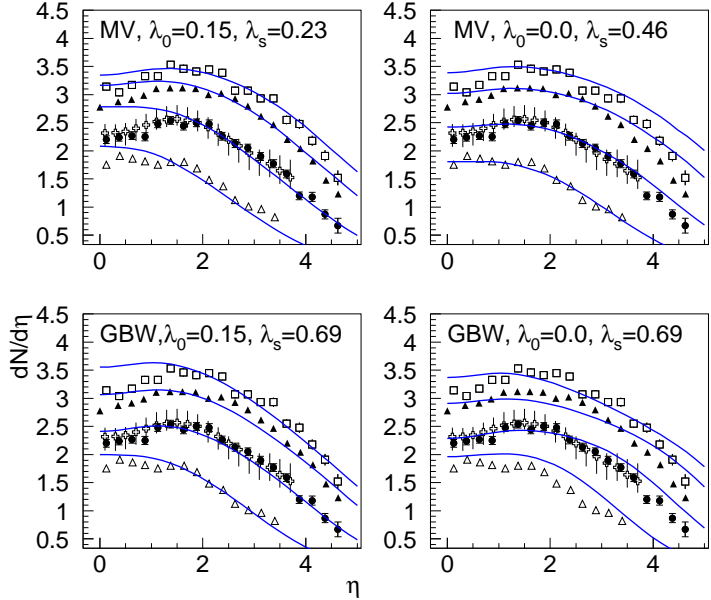}
\end{center}
\caption{ Pseudorapidity distributions $dN/d\eta$ for charged
particles from nucleon-nucleon collisions at UA5 energies~\cite{SPS}
$\sqrt{s}=53, 200, 546, 900 \, {\rm GeV}$ and PHOBOS
data~\cite{PHOBOS} at $\sqrt{s}=200 \, {\rm GeV}$. Upper plots:
initial distribution from the MV model, lower plots: initial
distribution from the GBW model. Left panels: $\lambda_0=0.15$, right
panels $\lambda_0=0.0$.  }
\label{fig:2}
\end{figure}
In figure~\ref{fig:2}, we plot the pseudo-rapidity distributions of
the charged particles produced in nucleon-nucleon collisions for
center of mass energies ranging from $53 \, {\rm GeV}$ to $900 \, {\rm
GeV}$.  The computations were performed for input distributions (for
BK evolution) at $x_0=0.01$ from the the GBW and MV models. The
normalization is a free parameter which is fitted at one energy.
Plots on the left of figure~\ref{fig:2} are obtained for
$\lambda_0=0.15$ (the free parameter in the large $x$ extrapolation)
whereas the right plots are for $\lambda_0=0.0$. The different values
of $\lambda_s=4.88 {\bar \alpha}_S$ are obtained for different values
of $\alpha_s$ as inputs to the BK equation. While these values of
$\alpha_s$ might appear small, they can be motivated as follows.  The
amplitude has the growth rate $\lambda_{\rm BK} = {2.77}\, \lambda_s /
{4.88} \approx 0.57\,\lambda_s$.  Thus $\lambda_s =0.46$, which gives
reasonable fits (more on this in the next paragraph) to the pp data
for the MV initial conditions, corresponds to $\lambda_{\rm BK} =
0.28$. Thus a small value of $\alpha_s$ in fixed coupling computations
``mimics'' the value for the energy dependence of the amplitude in
next-to-leading order resummed BK computations~\cite{Dionysis} and in
empirical dipole model comparisons to the HERA data~\cite{GBW}.

Our computations are extremely sensitive to the extrapolation
prescription to large $x$. This is not a surprise as the wave-function
of the projectile is probed at fairly large values of $x_1$. From our
analysis, we see that the data naively favors a non-zero value for
$\lambda_0$.  The value $\lambda_0=0$ results in distributions which,
in both the MV and GBW cases, give reasonable fits (albeit with
different normalizations) at lower energies but systematically become
harder relative to the data as the energy is increased.  To fit the
data in the MV model up to the highest UA5 energies, a lower value of
$\lambda_s$ than that in the GBW model is required. This is related to
the fact that MV model has tails which extend to larger values in
$k_\perp$ than in the GBW model. As the energy is increased, the
typical $k_\perp\sim Q_s(x_2)$ does as well. We will return to this
point shortly.
\begin{figure}[htbp]
\begin{center}
\includegraphics[width=4in]{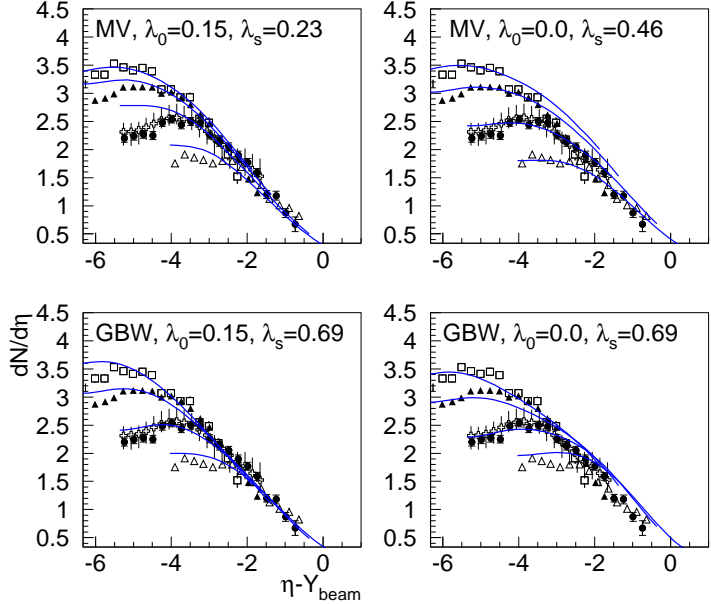}
\end{center}
\caption{ The same as figure~\ref{fig:2} but plotted versus
$\eta'=\eta-Y_{\rm beam}$ to illustrate the region of limiting
fragmentation.  }
\label{fig:3}
\end{figure}

In figure~\ref{fig:3} the same distributions are shown as a function
of the $\eta'=\eta-Y_{\rm beam}$.  The calculations for
$\lambda_0=0.15$ are consistent with scaling in the limiting
fragmentation region.  There is a slight discrepancy between the
calculations and the data in the mid-rapidity region. This discrepancy
may be a hint that one is seeing violations of $k_\perp$ factorization
in this regime because $k_\perp$ factorization becomes less reliable
the further one is from the dilute-dense kinematics of the
fragmentation regions~\cite{KrasnV4,Balitsky2}. This discrepancy
should grow with increasing energy. However, our parameters are not
sufficiently constrained that a conclusive statement can be made. For
instance, as we mentioned previously, there is a sensitivity to the
infrared mass chosen in the Jacobian of the transformation from $y$ to
$\eta$. This is discussed further in Ref.~\cite{GelisSV}.

\begin{figure}[htbp]
\begin{center}
\includegraphics[width=2.7in]{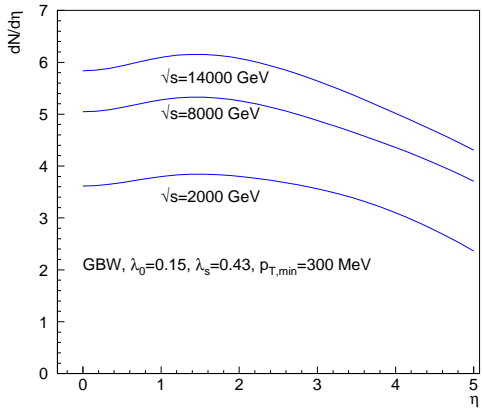}
\end{center}
\caption{ Predictions for higher center-of-mass energies for
proton-proton collisions: $\sqrt{s}=2,8,14 \, {\rm TeV}$ for GBW input
model. The parameter in the large $x$ extrapolation was set to
$\lambda_0=0.15$.}
\label{fig:4}
\end{figure}

In figure~\ref{fig:4} we show the extrapolation to higher energies, in
particular the LHC range of energies for the calculation with the GBW
input.  We observed previously that the MV initial distribution, when
evolved to these higher energies, gives a rapidity distribution which
is very flat in the range $-5<\eta<5$. We noted that this is because
the average transverse momentum grows with the energy giving a
significant contribution from the high $k_\perp$ tail of the
distribution in the MV input at $x_0$.  The effect of fragmentation
functions on softening the spectra in the limiting fragmentation
region can be simply understood by the following qualitative
argument. The inclusive hadron distribution can be expressed as
\begin{equation}
\frac{dN_h}{d^2 \p_\perp dy} 
= 
\int_{z_{\rm min}}^1 \frac{dz}{z}\, 
\frac{dN_g}{d^2 \q_\perp dy}\, D_{g\rightarrow h} \left(z=\frac{p_\perp}{q_\perp},\mu^2\right) \; ,
\label{eq:LIII-11}
\end{equation}
where $D_{g\rightarrow h}(z,\mu^2)$ is the fragmentation function
denoting the probability, at the scale $\mu^2$, that a gluon fragments
into a hadron carrying a fraction $z$ of its transverse momentum. For
simplicity, we only consider here the probability for gluons
fragmenting into the hadron. The lower limit of the integral can be
determined from the kinematic requirement that $x_{1,2}\leq 1$ -- we
obtain,
\begin{equation}
z_{\rm min} = \frac{q_\perp}{m_{\rm p}}\,e^{y-Y_{\rm beam}} \; .
\label{eq:LIII-12}
\end{equation}
If $z_{\rm min}$ were zero, the effect of including fragmentation
effects would simply be to multiply eq.~(\ref{eq:LIII-11}) by an overall
constant factor. At lower energies, the typical value of $q_\perp$ is
small for a fixed $y-Y_{\rm beam}$; the value of $z_{\rm min}$ is
quite low. However, as the center of mass energy is increased, the
typical $q_\perp$ value grows slowly with the energy. This has the
effect of raising $z_{\rm min}$ for a fixed $y-Y_{\rm beam}$, thereby
lowering the value of the multiplicity in eq.(\ref{eq:LIII-11}) for
that $y-Y_{\rm beam}$. Note further that eq.~(\ref{eq:LIII-12}) suggests
that there is a kinematic bound on $q_\perp$ as a function of
$y-Y_{\rm beam}$ -- only very soft gluons can contribute to the
inclusive multiplicity.

\begin{figure}[htbp]
\begin{center}
\includegraphics[width=2.7in]{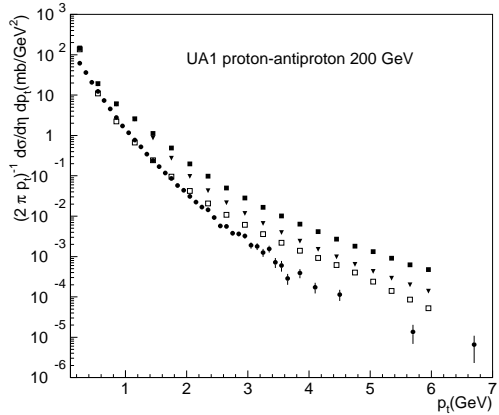}
\end{center}
\caption{ $p_\perp$ distribution from eq.~(\ref{eq:LIII-2}) with MV
(full squares) and GBW (full triangles) initial conditions.  The MV
initial condition -- with the fragmentation function included -- is
denoted by the open squares. The distribution is averaged over the
rapidity region $y=0.0-2.5$, to compare with data (in 200 GeV/nucleon
proton-antiproton collisions in the same pseudo-rapidity range) on
charged hadron $p_\perp$ distributions from the UA1 collaboration:
full circles.}
\label{fig:pt_pp}
\end{figure}

\begin{figure}[hbt]
\begin{center}
\includegraphics[width=4in]{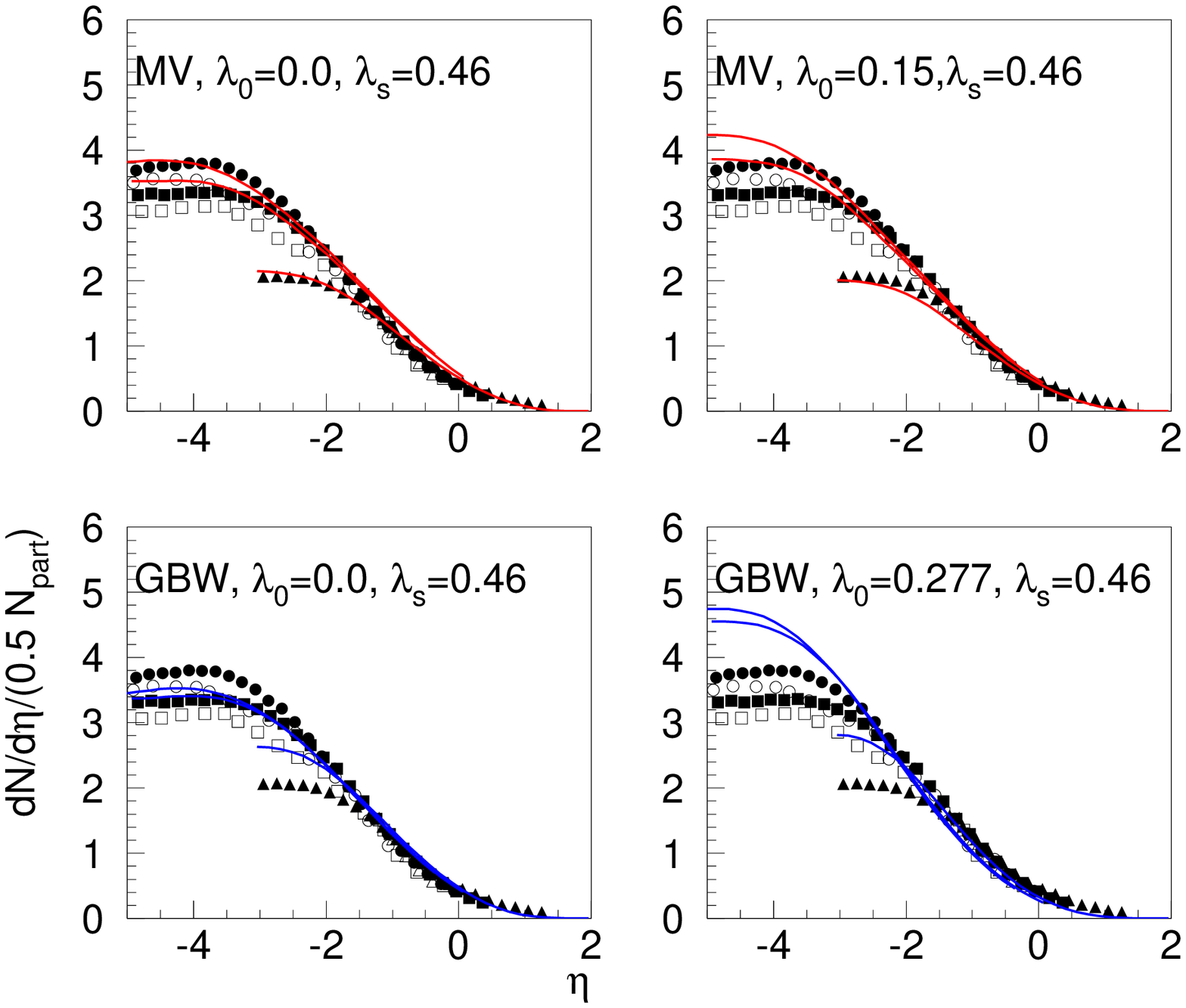}
\end{center}
\caption{ Pseudorapidity distributions normalized by the number of
participants for charged hadrons in gold-gold collisions from the
PHOBOS collaboration at energies $19.6, 130, 200 \, {\rm GeV}$ (filled
triangles , squares and circles), BRAHMS collaboration at energies
$130, 200 \, {\rm GeV}$ (open squares and circles) . The data from the
STAR collaboration at energy $62.4 \, {\rm GeV}$ (open triangles) are
not visible on this plot but can be seen more clearly in
fig.~\ref{fig:auauext}. Upper solid line: initial distributions from
the MV model; lower solid line: initial distributions from the GBW
model.  }
\label{fig:5}
\end{figure}

In figure~\ref{fig:pt_pp} we display the $p_\perp$ distributions
obtained from the MV input compare to the UA1 data \cite{UA1}.  We
compare the calculation with and without the fragmentation function.
The fragmentation function has been taken from \cite{KKPFRAG}.
Clearly the ``bare'' MV model does not describe the data at large
$k_\perp$ because it does not include fragmentation function effects
which, as discussed, make the spectrum steeper.  In contrast,
because the $k_\perp$ spectrum of the GBW model dies exponentially at
large $k_\perp$, this ``unphysical'' $k_\perp$ behavior mimics the
effect of fragmentation functions -- see figure \ref{fig:pt_pp}. Hence
extrapolations of this model, as shown in figure~\ref{fig:5} give a
more reasonable looking result. Similar conclusions were reached
previously in Ref.~\cite{Szczurek}.

\begin{figure}[htbp]
\begin{center}
\includegraphics[width=2.7in]{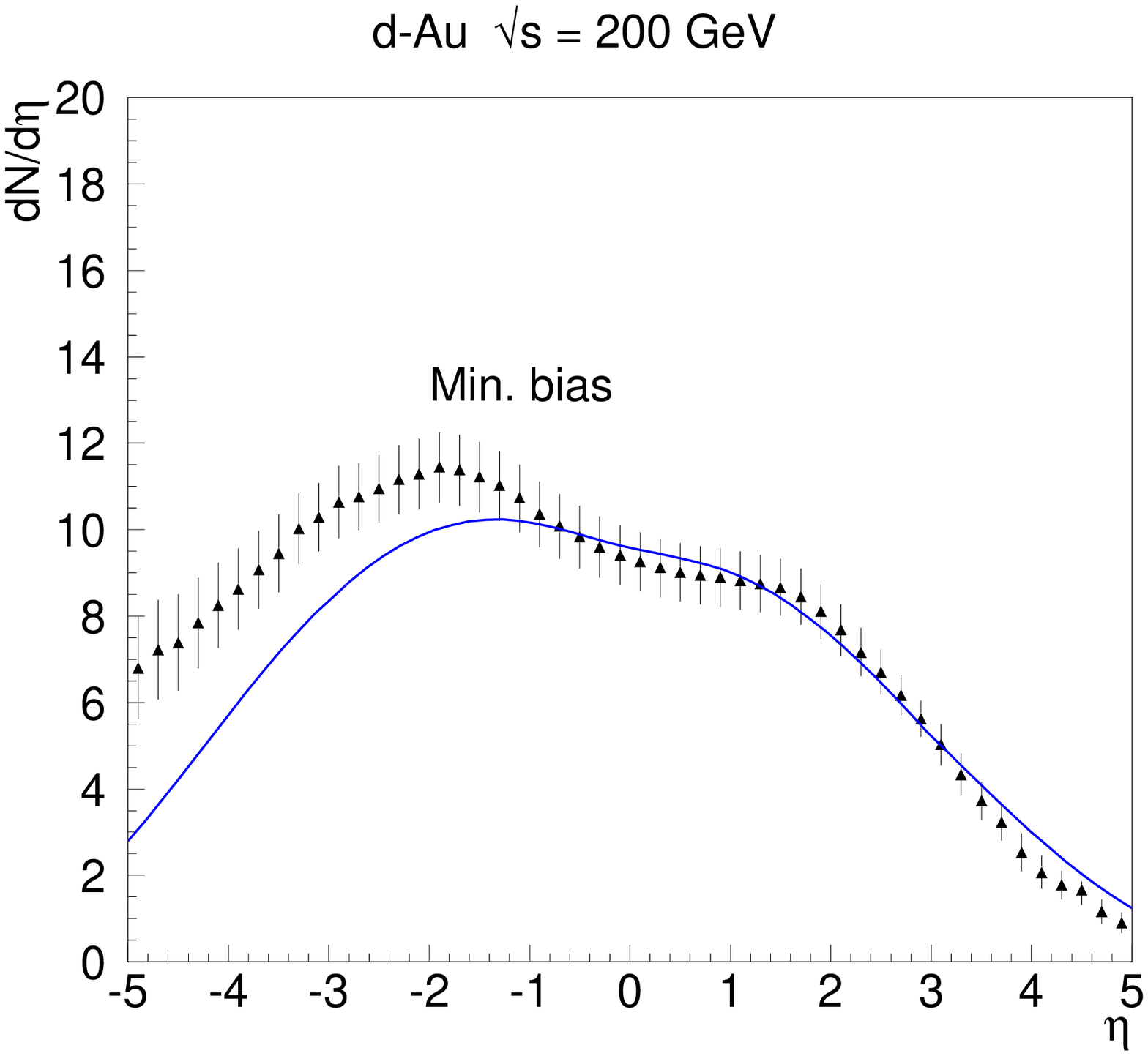}
\end{center}
\caption{Comparison of theory (with MV input) to minimum bias
deuteron-gold data at RHIC~\cite{PHOBOSDAU}.}
\label{fig:6}
\end{figure}

We next compute the pseudo-rapidity distribution in deuteron-gold
collisions.  In figure~\ref{fig:6} we show the result for the
calculation compared with the dA data~\cite{PHOBOSDAU}.  The unintegrated
gluons were extracted from the pp and AA data. The overall shape of
the distribution matches well on the deuteron side with the
minimum-bias data.  The disagreement on the nuclear fragmentation side
is easy to understand since, as mentioned earlier, it requires a
better implementation of nuclear geometry effects. Similar conclusions
were reached in Ref.~\cite{KLN1} in their comparisons to the RHIC
deuteron-gold data.

We now turn to to nucleus--nucleus collisions.  In figure~\ref{fig:5}
we present fits to data on the pseudo-rapidity distributions in
gold--gold collisions from the PHOBOS, BRAHMS and STAR collaborations.
The data~\cite{PHOBOS} are for $\sqrt{s} =19.6,130, 200 \, {\rm GeV}$
and the BRAHMS data~\cite{BRAHMS} are for $\sqrt{s}=130,200 \, {\rm
GeV}$.  A reasonable description of limiting fragmentation is achieved
in this case as well. One again has discrepancies in the central
rapidity region as in the pp case. We find that values of
$Q_s^A\approx 1.3$ GeV for the saturation scale give the best fits.
This value is consistent with the other estimates discussed
previously~\cite{KrasnNV1,KowalskiTeaney,KharzeevL}.  Apparently the
gold-gold data are better described by the calculations which have
$\lambda_0=0.0$. This might be related to the difference in the large
$x$ distributions in the proton and nucleus.  Further, slightly higher
values of $\lambda_s$ are preferred to the pp case.  This variation of
parameters from AA to pp case might be also connected with the fact
that in our approach the impact parameter is integrated out thereby
averaging over details of the nuclear geometry.

\begin{figure}[hbt]
\begin{center}
\includegraphics[width=2.7in]{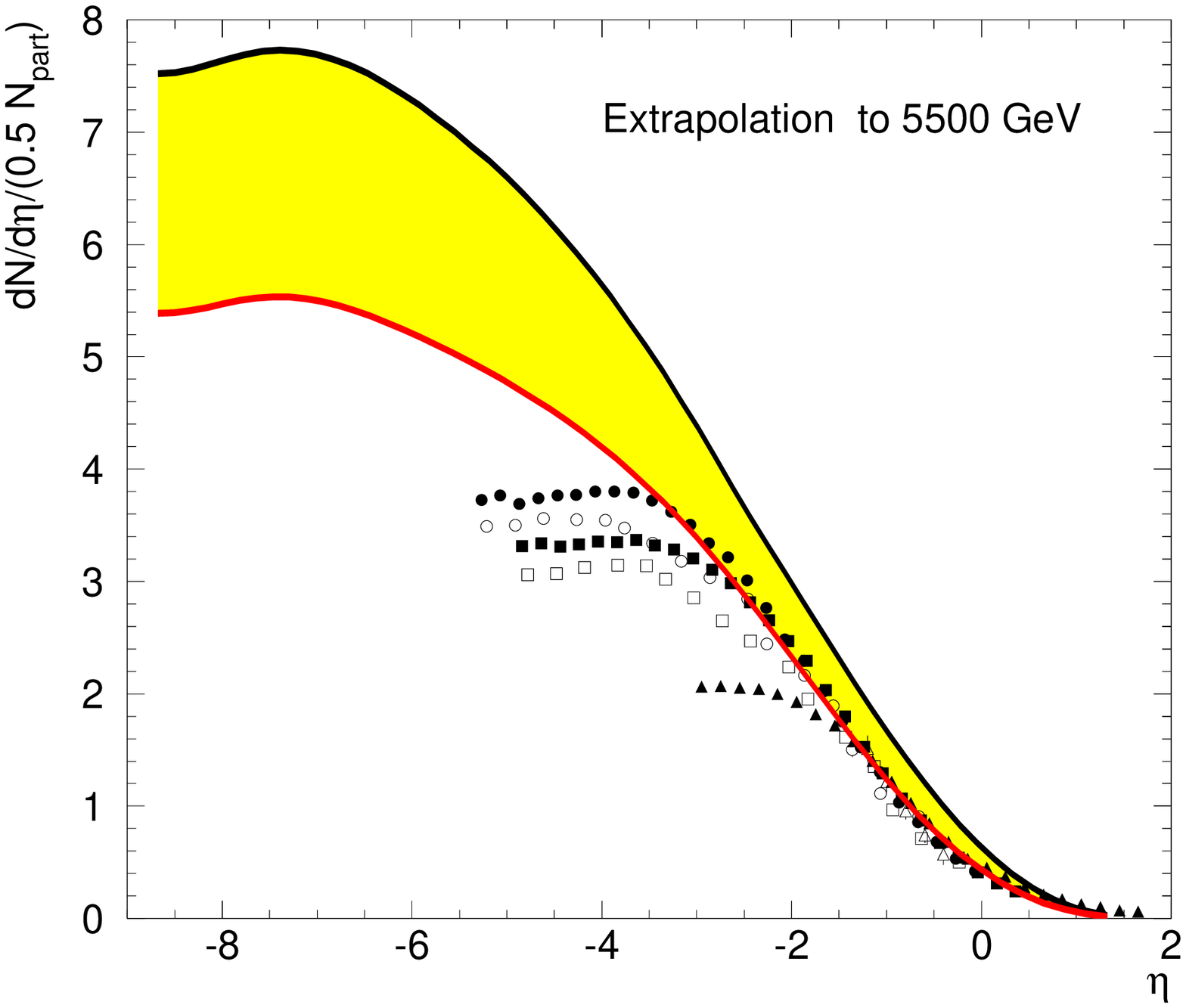}
\end{center}
\caption{ Extrapolation of calculations for gold-gold collisions shown
in fig.~\ref{fig:5} to the LHC energy $\sqrt{s}=5500 \; {\rm GeV}$/
nucleon. For comparison, the same data at lower energies are
shown. (See fig.~\ref{fig:5}.) Dashed line - GBW input
$\lambda_0=0,\lambda_s=0.46$, solid line - MV input with
$\lambda_0=0,\lambda_s=0.46$. }
\label{fig:auauext}
\end{figure}

In fig.~\ref{fig:auauext} we show the extrapolation of two
calculations to higher energy $\sqrt{s}=5500 \; {\rm GeV}$.  We note
that the calculation within the MV model gives results which would
violate the scaling in the limiting fragmentation region by
approximately $20\%$ at larger $y-Y_{\rm beam}$.  This violation is
due partly to the effect of fragmentation functions discussed
previously and partly to the fact that the integrated parton
distributions from the MV model do not obey Bjorken scaling at large
values of $x$.  In the latter case, the violations are proportional to
$\ln(Q_s^2(x_2))$ as discussed previously. The effects of the former
are simulated by the GBW model -- the extrapolation of which, to
higher energies, is shown by the dashed line. The band separating the
two therefore suggests the systematic uncertainity in such an
extrapolation coming from (i) the choice of initial conditions and
(ii) the effects of fragmentation functions which are also uncertain
at lower transverse momenta.

To summarize the discussion in this lecture, we studied the phenomenon
of limiting fragmentation in the Color Glass Condensate framework. In
the dilute-dense (projectile-target) kinematics of the fragmentation
regions, one can derive (in this framework) an expression for
inclusive gluon distributions which is $k_\perp$ factorizable into the
product of ``unintegrated'' gluon distributions in the projectile and
target.  From the general formula for gluon production
(eq.~(\ref{eq:LIII-2})), limiting fragmentation is a consequence of
two factors:
\begin{itemize}
\item Unitarity of the $U$ matrices which appear in the definition of
the unintegrated gluon distribution in eq.~(\ref{eq:LIII-4}).
\item Bjorken scaling at large $x_1$, namely, the fact that the
integrated gluon distribution at large $x$, depends only on $x_1$ and
not on the scale $Q_s(x_2)$. (The residual scale dependence
consequently leads to the dependence on the total center-of-mass
energy.)
\end{itemize}  
Deviations from the limiting fragmentation curve at experimentally
accessible energies are very interesting because they can potentially
teach us about how parton distributions evolve at high energies. In
the CGC framework, the Balitsky-Kovchegov equation determines the
evolution of the unintegrated parton distributions with energy from an
initial scale in $x$ chosen here to be $x_0=0.01$. This choice of
scale is inspired by model comparisons to the HERA data.

We compared our results to data on limiting fragmentation from pp
collisions at various experimental facilities over a wide range of
collider energies, and to collider data from RHIC for deuteron-gold
and gold-gold collisions. We obtained results for two different models
of initial conditions at $x\geq x_0$; the McLerran-Venugopalan model
(MV) and the Golec-Biernat--Wusthoff (GBW) model.  

We found reasonable agreement for this wide range of collider data for
the limited set of parameters and made predictions which can be tested
in proton-proton and nucleus-nucleus collisions at the LHC. Clearly
these results can be fine tuned by introducing further details about
nuclear geometry.  More parameters are introduced, however there is
more data for different centrality cuts -- we leave these detailed
comparisons for future studies. In addition, an important effect,
which improves agreement with data, is to account for the
fragmentation of gluons in hadrons. In particular, the MV model, which
has the right leading order large $k_\perp$ behavior at the partonic
level, but no fragmentation effects, is much harder than the data. The
latter falls as a much higher power of $k_\perp$. As rapidity
distributions at higher energies are more sensitive to larger
$k_\perp$, we expect this discrepancy to show up in our studies of
limiting fragmentation and indeed it does. Taking this
into account leads to more plausible extrapolations of fits of
existing data to LHC energies.

\section*{Acknowledgments}
These lectures were delivered by one of us (RV) at the 46th Zakopane
school in theoretical physics. This school held a special significance
because it coincided with the 70th birthday of Andrzej Bialas who
is a founding member of the school. RV would like to thank Michal
Praszalowicz for his excellent organization of the school. This work
has drawn on recent results obtained by one or both of us in collaboration
with H. Fujii, K. Fukushima, S. Jeon, K. Kajantie, T. Lappi, L.
McLerran, P. Romatschke and A. Stasto. We thank them all. RV was supported by DOE
Contract No. DE-AC02-98CH10886.

\end{document}